\documentclass[sigconf, nonacm]{acmart}

\newcommand\vldbdoi{XX.XX/XXX.XX}

\newcommand\vldbavailabilityurl{https://github.com/danielxue/global-hash-tables-strike-back}
\newcommand\vldbpagestyle{plain}

\usepackage{algorithm,algpseudocode}
\usepackage{float}
\newfloat{algorithm}{t}{lop}

\usepackage{balance}
\usepackage{soul}
\usepackage{pifont}
\newcommand{\circleOne}{\ding{182}\xspace}
\newcommand{\circleTwo}{\ding{183}\xspace}
\newcommand{\circleThree}{\ding{184}\xspace}

\usepackage[dvipsnames]{xcolor}
\usepackage{xspace}

\usepackage{setspace}

\begin{document}
\title{Global Hash Tables Strike Back! An Analysis of Parallel GROUP BY Aggregation}

\author{Daniel Xue}
\orcid{0009-0007-6892-8437}
\affiliation{\institution{University of Pennsylvania}}
\email{danxue@seas.upenn.edu}

\author{Ryan Marcus}
\orcid{0000-0002-1279-1124}
\affiliation{\institution{University of Pennsylvania}}
\email{rcmarcus@seas.upenn.edu}

\begin{abstract}
Efficiently computing group aggregations (i.e., \texttt{GROUP BY}) on modern architectures is critical for analytic database systems. Hash-based approaches in today's engines predominantly use a partitioned approach, in which incoming data is partitioned by key values so that every row for a particular key is sent to the same thread. In this paper, we revisit a simpler strategy: a fully concurrent aggregation technique using a shared hash table. While approaches using general-purpose concurrent hash tables have generally been found to perform worse than partitioning-based approaches, we argue that the key ingredient is \emph{customizing the concurrent hash table for the specific task of group aggregation}. Through experiments on synthetic workloads (varying key cardinality, skew, and thread count), we demonstrate that in morsel-driven systems, a purpose-built concurrent hash table can match or surpass partitioning-based techniques. We also analyze the operational characteristics of both techniques, including resizing costs and memory pressure. In the process, we derive practical guidelines for database implementers. Overall, our analysis indicates that fully concurrent group aggregation is a viable alternative to partitioning. 
\end{abstract}

\maketitle

\pagestyle{\vldbpagestyle}

\ifdefempty{\vldbavailabilityurl}{}{
\vspace{.3cm}
\begingroup\small\noindent\raggedright\textbf{PVLDB Artifact Availability:}\\
The source code, data, and/or other artifacts have been made available at \url{\vldbavailabilityurl}.
\endgroup
}
\setcounter{page}{1} 
\balance 

\section{Introduction}

Group aggregation, represented in SQL via \texttt{GROUP BY}, is a fundamental operation in analytical query processing, especially decision-support workloads~\cite{nambiar_making_2006}. To ensure that database systems continue to scale well with new many-core architectures, it is critical to build highly concurrent group aggregation schemes. While many different schemes for group aggregation exist (e.g., sorting), this paper will analyze hash-based aggregation techniques.

Analytic database systems today are quite diverse. For example, Datafusion~\cite{lamb_aggregating_2023} follows a Volcano-style~\cite{graefe_volcano_1993} block iteration approach (i.e., ``pull''), whereas DuckDB~\cite{raasveldt_duckdb_2019} follows a HyPer-inspired morsel-driven parallelism~\cite{leis_morsel-driven_2014} approach (i.e., ``push''). Despite drastic differences in their execution models, nearly all of today's analytic database systems use partitioning techniques to parallelize group aggregations. To the best of our knowledge, the partitioning approach proposed by Raman et al. ~\cite{raman_db2_2013} has become dominant in many modern analytic systems~\cite{leis_morsel-driven_2014, kuiper_robust_2024, lamb_aggregating_2023}, which uses preaggregation to alleviate pressure from data skew. 
We provide background on partitioning-based techniques in~\autoref{sec:part_agg}.

An alternative to partition-based approaches is to use a global concurrent hash table. Instead of partitioning keys into groups, each worker can concurrently access a global hash table. In theory, such a hash table has many operational benefits, such as lower memory usage, reducing the impact of skew, and simplifying implementations. In practice, despite several improvements to general-purpose concurrent hash tables~\cite{li_algorithmic_2014, pandey_iceberght_2023, maier_concurrent_2018}, contention effects and synchronization overhead represent significant scalability barriers.

But is it really surprising that general-purpose concurrent hash tables perform worse than purpose-built solutions like partitioned group aggregation? After all, general-purpose hash tables must support a myriad of operations that are irrelevant to group aggregation, such as deletes and shrinking. General-purpose hash tables must also be optimized for a wide range of workloads, where deletes, inserts, and lookups might come from different threads, in different distributions, and at different times. \ul{A concurrent hash table optimized for group aggregation could sidestep most of this complexity}: the only required operation is the aggregation of a new value, and it is reasonable to assume that every thread will invoke this operation consistently until all data is consumed.

\subsubsection*{\normalfont\bfseries Main result}
In this paper, we explore the design space of group aggregation algorithms using a global concurrent hash table, comparing against the state-of-the-art partitioned approach in morsel-driven~\cite{leis_morsel-driven_2014} execution engines. Most significantly, \ul{we find that a simple, purpose-built concurrent hash table using linear probing and a customized get-or-insert function can scale well on modern multi-core hardware and match or even outperform partitioning-based approaches.} Our results do not show that one approach is better than the other, but instead highlight the operational benefits and costs of both approaches. Throughout our exploration, we make specific recommendations for database implementers.

Our implementation of group aggregation with a global concurrent hash table closely tracks the implementation in MonetDB~\cite{boncz_monetdbx100_2005}. Each worker, upon receiving a row, first obtains from a hash table an integer ``ticket'' for that row's grouping key. This ticket uniquely identifies each group and serves as an index to locate the aggregated value for that group. We provide a detailed description of this procedure in ~\autoref{sec:fully_concurrent_expl}. This two-phase procedure, ticketing and partial aggregate update, is repeated for each row, and opens up a large number of possible designs, which we explore.

\subsubsection*{\normalfont\bfseries Ticketing}
In the initial phase of fully concurrent group aggregation, each unique group is assigned an integer ``ticket.'' This operation can be performed with a concurrent hash table that atomically checks if a key is already in the table, returning the ticket for that key if so, and, if not, inserting a new ticket into the table. Surprisingly, many general-purpose hash tables cannot perform this operation atomically or do not optimize for this particular case. As a result, there is significant room for improvement by building a specialized fast path for this particular operation, while avoiding adding extra overhead for unnecessary operations like deletes. We test several implementations based on atomics and fine-grained locking, and show that \ul{simple purpose-built hash tables can significantly outperform their complex general-purpose counterparts.}

\subsubsection*{\normalfont\bfseries Partial Aggregate Update}
After ticketing, each worker must apply the relevant aggregation function on the values associated with each row. This can be done in either a thread-specific way (e.g., each thread maintains local aggregation storage, merging the results at the end), or using concurrent access to global space. We explore the tradeoffs between both approaches, characterizing their operational tradeoffs. Specifically, we find that concurrent access to global space works well in the absence of heavy hitters, and \ul{we propose a simple thread local approach that works well except for when every grouping key is unique.}

\subsubsection*{\normalfont\bfseries Organization}
Our experimental study is organized as follows. 
In~\autoref{sec:prelim}, we explain our assumed model of execution, and introduce the basics of partitioned aggregation and concurrent aggregation. We also introduce the experimental setup used throughout our experiments. 
In~\autoref{sec:fca}, we investigate the design space of aggregation with a concurrent hash table. In~\autoref{sec:ticketing}, we explore the ticketing step, and in~\autoref{sec:update}, we explore the aggregation step.
In~\autoref{sec:experiments}, we analyze the end-to-end performance of both approaches. 
In~\autoref{sec:related} we discuss related works, before concluding and outlining future work in~\autoref{sec:conclusion}.

\section{Preliminaries} \label{sec:prelim}
In this section, we first establish the model of query execution we are operating within and discuss the relevant constraints it places upon our work. We then provide an overview of the two models of aggregation assesed in this paper: partitioned and fully concurrent. 

\subsection{Model of Execution}
Leis et al.~\cite{leis_morsel-driven_2014} introduced the morsel-driven framework for parallel execution to improve performance in main-memory systems, where latency is compute-bound rather than I/O-bound. In this model, parallelism is achieved by breaking down work into data fragments called ``morsels.'' Morsels are dynamically assigned to threads in a pool, using work-stealing to ensure even distribution. A thread carries a morsel through the entire pipeline before grabbing another. Morsel-driven parallelism is often used with columnar data representations (e.g., HyPer~\cite{leis_morsel-driven_2014}), with morsels serving as units for vector-at-a-time execution. Operating on a dense vector of values enables important optimizations, such as amortizing interpretation overhead and SIMD instructions~\cite{kersten_everything_2018}. However, the morsel-driven model places some restrictions on operator implementations. Operators must be able to operate on chunks of data at a time to conform with pipelining and cannot access the rest of the incoming tuples. 

In this work, we focus on the morsel-driven execution model due to its popularity in open source OLAP systems (e.g.,~\cite{raasveldt_duckdb_2019}). While our conclusions and recommendations may apply to other execution models (e.g., Datafusion~\cite{lamb_aggregating_2023} uses operator-level parallelism but adopts an aggregation strategy from a morsel-driven system), different execution models may also lead to significant changes in performance properties. We leave investigations under alternative execution models to future work.

\subsection{Partitioned Aggregation Model} \label{sec:part_agg}
In a partitioned approach, synchronization is avoided by splitting the key domain among threads. A naive partitioning strategy assigns incoming tuples to their proper thread (typically by taking the radix or mod of the key's hash), which then aggregates the assigned tuples locally. However, this method suffers significantly from data skew. A simple solution is to use a local hash table to preaggregate heavy hitters before partitioning~\cite{ye_scalable_2011}. 

\begin{figure}
  \centering
  \includegraphics[width=\linewidth]{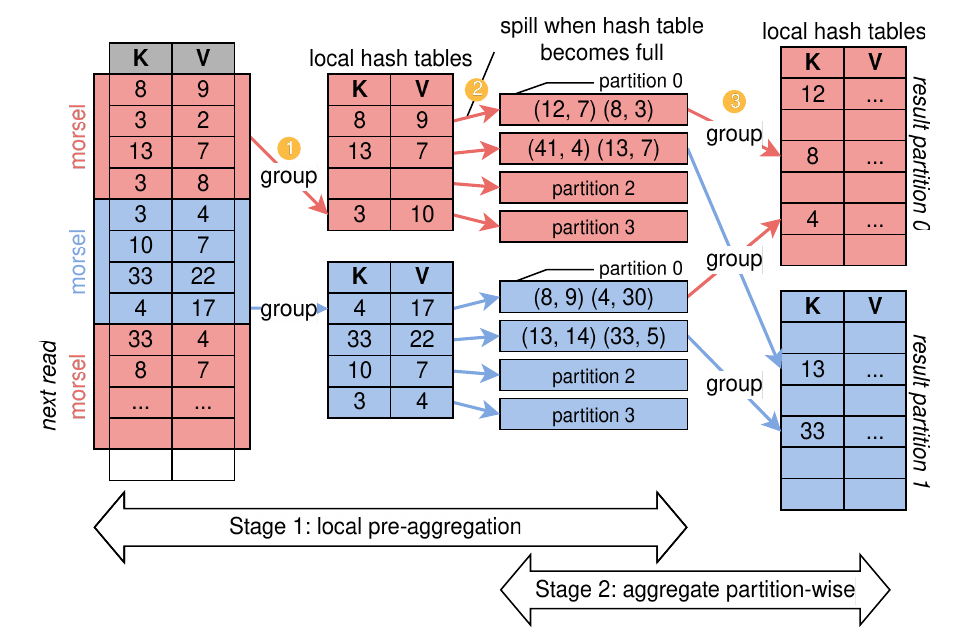}
  \Description{A sample execution of partitioned aggregation. This diagram is modified slightly from the one presented in~\cite{leis_morsel-driven_2014} and is provided to contrast with the sample execution by our fully concurrent model on the same input in~\autoref{fig:concurrent_aggregation_stages}.}
  \caption{A sample execution of partitioned aggregation. This diagram is modified slightly from the one presented in~\cite{leis_morsel-driven_2014} and is provided to contrast with the sample execution by our fully concurrent model on the same input in~\autoref{fig:concurrent_aggregation_stages}.}
  \label{fig:partitioned_aggregation_stages}
\end{figure}

Leis et al.~\cite{leis_morsel-driven_2014} use a partitioned aggregation strategy (originating in~\cite{raman_db2_2013}) illustrated in~\autoref{fig:partitioned_aggregation_stages}. This method consists of two stages: local preaggregation and partition-wise aggregation. First, each thread aggregates all values from their assigned morsel(s) in a fixed-sized local hash table \textcolor{Dandelion}{\circleOne}. When the hash tables are full, the partially aggregated values are spilled into partitions \textcolor{Dandelion}{\circleTwo} and the process continues. In the second stage, partitions are exchanged between threads, and each thread combines all the partial aggregates to compute the final answer \textcolor{Dandelion}{\circleThree}.

This algorithm is motivated by the need for skew resistance, avoiding uneven work distribution by spreading the work of aggregating heavy hitters among all threads in the local preaggregation stage. However, for high cardinality workloads, there is repeated spilling from the local aggregation tables. The constant spilling results in each tuple essentially being aggregated twice, once in each stage of the algorithm, creating a significant source of overhead. 

The local preaggregation approach to partitioned aggregation has been adopted by several real-world systems, including DuckDB~\cite{kuiper_robust_2024} and Datafusion~\cite{lamb_aggregating_2023}. Due to the widespread adoption of this particular algorithm and its good scaling behavior~\cite{ye_scalable_2011, kuiper_robust_2024}, it is used here as the baseline partitioned aggregation method. 

\subsection{Fully Concurrent Aggregation Model} \label{sec:fully_concurrent_expl}
\begin{figure}
  \centering
  \includegraphics[width=\linewidth]{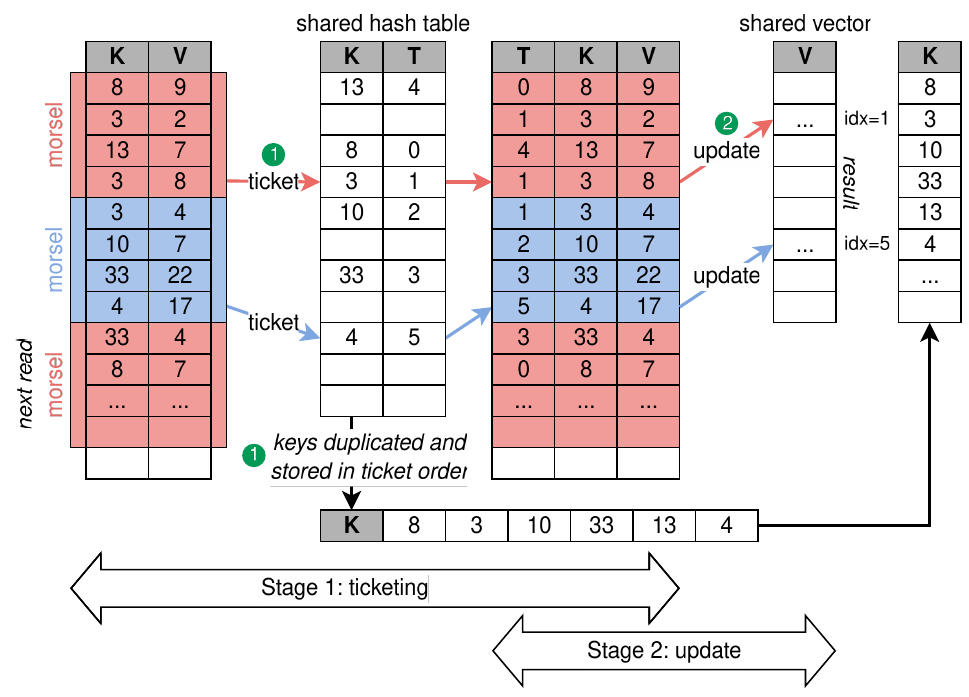}
  \Description{A sample execution of our fully concurrent aggregation model using the same instance as in~\autoref{fig:partitioned_aggregation_stages}. This particular diagram aligns most closely to our atomic or locking method of partial aggregate updates. In contrast to partitioned aggregation, instead of completing all work in local data structures, there are now two shared structures, the ticketing hash table and the vector of partial aggregates.}
  \caption{A sample execution of our fully concurrent aggregation model using the same instance as in~\autoref{fig:partitioned_aggregation_stages}.}
  \label{fig:concurrent_aggregation_stages}
\end{figure}

To perform a group by aggregation in a fully concurrent manner, each thread must aggregate an arbitrary morsel of data from start to finish, without any key distribution. We separate fully concurrent aggregation into two different steps: \emph{ticketing} and partial aggregation \emph{update}, as shown in~\autoref{fig:concurrent_aggregation_stages}. 

During ticketing \textcolor{ForestGreen}{\circleOne}, a concurrent shared hash table is used to map each key value to an integer ``ticket.'' This mapping is one-to-one: each unique key is granted a single, unique ticket, and the ticket assigned to a key is consistent across all threads due to the use of a shared hash table. A ticket conceptually represents the location where the corresponding partial aggregate to be updated is stored, which in our implementation is simply an index of a vector. Alternatively, the ticketing step can be viewed as incrementally creating a perfect hash function. When we insert a new key, we also duplicate the key and store it in a vector in ticket order, which is later used to reconstruct the key column.

As a result of the first step, we have a morsel-sized column of tickets, stored as a vector. This column of tickets (T) and the input morsel of values (V) are fed to the update step \textcolor{ForestGreen}{\circleTwo}, where the query's specified aggregation function(s) are applied (e.g., incremented for \texttt{COUNT}, added for \texttt{SUM}, etc). This step is the ``aggregation'' part of group by aggregation. 

\subsubsection*{\normalfont\bfseries Ticketing Indirection} 
Note that this layer of indirection that separates the partial aggregate from the hash map is already commonplace in practice. For example, it is used by MonetDB ~\cite{boncz_monetdbx100_2005} to enable vectorized execution, as well as by DuckDB~\cite{kuiper_robust_2024} and DataFusion~\cite{lamb_aggregating_2023}. Indirection also enables optimizations only possible when acting on a dense column of values, such as SIMD, or to amortize the cost of dynamic type resolution in non-compiled systems. 

Given that indirection is already commonplace, we formalize it in our model to take advantage of its logical properties. In particular, since each unique key is assigned a ticket exactly once, our hash table workload is reduced to only lookups and inserts. We take advantage of this fact by designing a customized concurrent hash table that supports only these operations. This indirection also pushes concurrent updates to the second stage. Thus, concurrent updates act on a vector of values, rather than units of a hash table, allowing for alternate ways of synchronizing writes that do not degrade the performance of the ticketing hash table. 

\subsection{Experimental setup} \label{sec:setup}
Since our analysis and discussion will integrate several microexperiments, we first describe our experimental setup. Readers may wish to skip this section and reference it as needed.

We primarily run our experiments on a machine with 256 GB of RAM and an AMD EPYC 9454P processor with 48 cores @2.75GHz (``AMD''). This chip supports simultaneous multi-threading (SMT) up to 96 threads, has 3MB L1 cache, 48MB L2 cache, and a 256MB shared L3 cache. To test a diversity of architectures, we also run tests on an ARM system with an Ampere Altra Q80-30 @3.0GHz with 80 cores (no SMT support), 256GB RAM, 10MB L1 cache, and 80MB L2 cache (``ARM''), and an Intel system with an Intel Xeon Gold 5412U @2.10GHz with 24 cores (SMT up to 48 threads), 256GB RAM, 2MB L1 cache, 48MB L2 cache, 45MB shared L3 cache (``Intel''). For all platforms, clock boosting was disabled, as was address space layout randomization (which caused significant performance variation from run to run). All experiments were implemented in Rust and were compiled in release mode using the \texttt{target-cpu=native} flag.

\subsubsection*{\normalfont\bfseries Datasets}
Our tests use synthetic datasets consisting of 100 million key-value pairs. Each key and value is a 64-bit integer (other than in~\autoref{sec:tuple_size}). We varied the cardinality of the dataset, labeled low, high, and unique, which consist of 1000 unique keys (100\% lookup/0\% insertion on ticketing), 10 million unique keys (90\% lookup/10\% insertion), and 100 million unique keys (0\% lookup/100\% insertion) respectively, distributed perfectly uniformly. On the high cardinality dataset, we also add two types of skew: a Zipfian distribution with exponent parameter $s=0.8$, as well as a heavy-hitter dataset where 50\% of the dataset consists of the same key. 

\subsubsection*{\normalfont\bfseries Workload}
As a demonstrative example of common aggregation functions, we use \texttt{SUM} for all tests. All experiments (except those in~\autoref{sec:resizing}) assume perfect cardinality estimation and thus perfectly size hash tables and partial aggregate vectors. All results are obtained by taking the median latency of a given workload after 9 runs (not including warm-up runs).

\section{Fully Concurrent Aggregation} \label{sec:fca}
Here, we investigate the design space for each stage of fully concurrent aggregation. We discuss methods for and execute micro-benchmarks of each stage in isolation to identify promising methods, which we later test end-to-end in~\autoref{sec:experiments}. 

\subsection{Ticketing} \label{sec:ticketing}
As described in~\autoref{sec:prelim}, the ticketing step is performed using a concurrent hash table to map each unique key to a unique and immutable ``ticket.'' In this section, we first establish the interface of hash tables designed for ticketing and the process used to generate tickets. Then, we discuss various candidate hash table designs, followed by experimental evaluation and discussion. 

\subsubsection*{\normalfont\bfseries Interface} When a new key arrives, an insert operation is required. Following a successful insertion of a key, all future requests must look up the inserted ticket value. In many concurrent hash table designs, this lookup operation can be achieved with much less overhead than an insert (e.g., shared locks are cheaper than exclusive locks). However, since we do not know ahead of time whether an insert is necessary (i.e., we do not know if a particular key has already been given a ticket), it is crucial to provide an efficient fast path for an atomic get-or-insert operation. Such an operation is rarely supported out-of-the-box by concurrent hash maps, but depending on the specific hash map implementation, the same effect could be achieved using an entry API or with a lookup followed by a non-overwriting insert. 

To efficiently retrieve keys in the same order as our aggregate values when materializing our final results, we also need the hash table to store a copy of keys in ticket order. This creates some performance overhead but is generally efficient since there is no contention--only the thread issuing a ticket is responsible for storing a copy. The ticket-order copy of keys can also be maintained as the only copy to decrease memory overhead (e.g., since the ticket value can be used to lookup the key during hash map operations), but would come at the cost of an additional cache miss per lookup. 

\begin{algorithm}
    \caption{Get-or-insert implementation for Folklore*.}
    \begin{algorithmic}
        \Function{GET\_OR\_INSERT}{key, ticket, table}
            \State $idx := key.hash() \bmod table.len()$
            \Loop \Comment Atomic operations use acquire/release ordering.
                \State $(t, k) := (table[idx].t.load(), table.[idx].k.load())$
                \If{$t \geq 2$} \Comment Fast path lookup.
                    \If{$k=key$}
                        \State \Return $t - 2$
                    \Else
                        \State $idx := (idx + 1) \bmod table.len()$
                    \EndIf
                \ElsIf{$table[idx].t.CAS(0, 1)$} \Comment Insert.
                    \State $table[idx].k.store(key)$
                    \State $table[idx].t.store(ticket + 2)$
                    \State \Return $ticket$
                \EndIf
            \EndLoop
        \EndFunction
    \end{algorithmic}
    \label{alg:get_or_insert}
\end{algorithm}

\subsubsection*{\normalfont\bfseries Hash Table Designs} We benchmark a variety of state-of-the-art hash table designs. These include cuckoo hashing~\cite{li_algorithmic_2014}, Iceberg hashing~\cite{bender_iceberg_2023, pandey_iceberght_2023}, Rust's Leapfrog~\cite{clucas_leapfrog_nodate} library, which uses leapfrog probing~\cite{preshing_leapfrog_2016}, and Rust's popular DashMap~\cite{wejdenstal_dashmap_nodate}.

We also implement a variant of the Folklore hash table proposed by Maier et al.~\cite{maier_concurrent_2018}, a lockless linear probing hash table. Our implementation leverages that we control the value being stored in the table (i.e., the ticket). We reserve 0 to indicate the key-value pair of the slot is empty and 1 to indicate a write is in progress. These properties allow our design to use only a single-word \texttt{CAS} instruction instead of the two-word version (which is not universally supported) required by the canonical implementation. We denote our implementation Folklore* to indicate the differences. The \texttt{GET\_OR\_INSERT} procedure for this design is outlined in~\autoref{alg:get_or_insert}. 

Our implementations additionally benefit from vectorized execution that optimizes hashing and amortizes the acquisition of a read lock on the shared table (which is needed to ensure the correctness of resizing). 

\begin{figure}
  \centering
  \includegraphics[width=\linewidth]{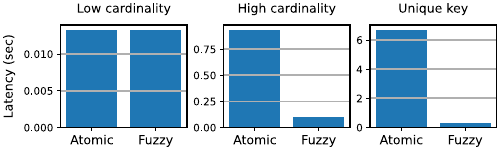}
  \Description{Performance of the Folklore* hash table when using a fuzzy ticketer as opposed to an incrementing atomic counter to issue tickets. Performance is evaluated with both low and high cardinality workloads and is measured with latency (lower is better).}
  \caption{Performance of the Folklore* hash table with a fuzzy ticketer as opposed to an incrementing atomic counter. Performance is evaluated with latency (lower is better).}
  \label{fig:fuzzy_ticketer}
\end{figure}

\subsubsection*{\normalfont\bfseries Micro Experiment: Generating Ticket Values} Generating a ticket value is simple on the surface, but actually presents a potential source of contention. Multiple threads must avoid issuing the same ticket to different keys. A naive implementation of a ``ticketer'' would use an atomic counter, incremented whenever a thread needs a new ticket. However, contention on the counter is high for insert-heavy workloads. To combat this issue, one can use a \emph{fuzzy ticketer} that assigns each thread a range of tickets to issue at a time. Each thread only needs to concurrently access an atomic value when it has exhausted its assigned range. The tradeoff is that the vector of partial aggregates may no longer be perfectly dense, but the number of gaps is bounded linearly by the number of threads and does not cause much overhead to remove.

There is a large latency gap on insert-heavy workloads between using a pure atomic counter and our fuzzy ticketer, as seen in~\autoref{fig:fuzzy_ticketer} (lower latency is better). This micro-benchmark is tested on our best-performing hash table design, Folklore* at 48 threads. Latency in the high cardinality workload is greater for the pure atomic method by a factor of 9.0x, and even worse for unique keys. Since using a single atomic ticket value degrades performance when there are many unique keys, \ul{we recommend that implementers amortizes the cost of concurrent accesses across multiple inserts,} such as with our fuzzy ticketer. 

\subsubsection*{\normalfont\bfseries Micro Experiment: Memory Allocation} \label{sec:opt}
For memory-intensive database workloads such high-cardinality aggregations, the performance of the memory allocator can have a notable impact on the overall latency: a global hash table must be allocated. Prior work has compared various memory allocators based on their performance, memory efficiency, and scalability~\cite{durner_impact_2019}. The allocation of a large global hash table for high cardinalities is not only time-consuming, but is crucially also single-threaded. This single-threaded work is a constant cost that does not scale with threads, and can severely degrade performance at high thread counts. In~\autoref{tab:zero_alloc}, at 24 threads we can see that pre-optimization memory allocation can be more than half of total latency on some systems.

\begin{table}
  \caption{Impact of zero-allocation on Folklore* performance on AMD and Intel at 24 threads with unique keys.}
  \label{tab:zero_alloc}
  \begin{tabular}{ccccc}
    \toprule
    System & \multicolumn{2}{c}{AMD} & \multicolumn{2}{c}{Intel} \\
    Zero-allocate & No & Yes & No & Yes \\
    \midrule
    Total latency (sec) & 0.48 & 0.38 & 0.92 & 0.27 \\
    Allocation (sec) & 0.24 & 0.00 & 0.68 & 0.00 \\
    Ticketing (ec) & 0.23 & 0.37 & 0.23 & 0.27 \\
    \bottomrule
\end{tabular}
\end{table}

Zero-allocation was used to optimize the memory allocation step and improving the scaling of concurrent hash tables. Although zero-allocation is restricted to only certain types, this restriction is not a problem for the ticketing use case as long as the hash table is designed to not prematurely access uninitialized key fields. Zero-allocation, often a special type of call to the allocator (i.e., \texttt{calloc} in C), enables a number of optimizations by the operating system. A particularly important one is copy-on-write optimization, which defers allocation of a page to its first access rather than on the initial call to the allocator. As a result, memory allocation is no longer on the single-threaded critical path. When using zero-allocation, we see in~\autoref{tab:zero_alloc} that the initial allocation becomes negligible in exchange for a modest performance hit during the actual ticketing step. \ul{Thus, we encourage practitioners to consider zeroed memory allocation strategies for large tables.}

\begin{figure*}
  \centering
  \includegraphics[width=\linewidth]{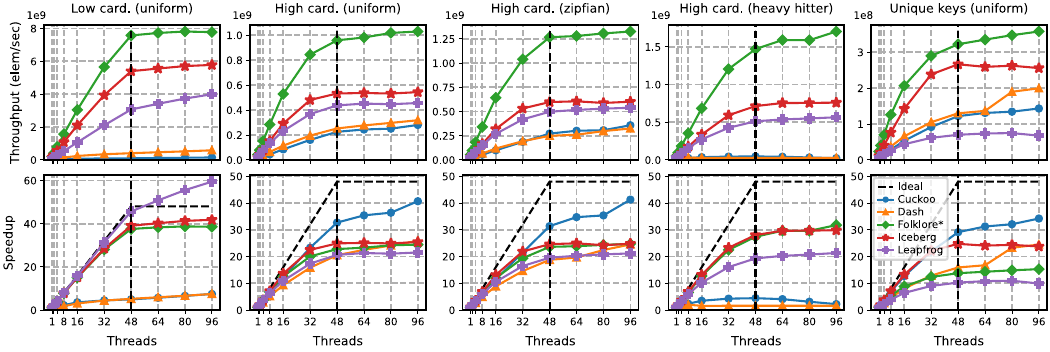}
  \Description{Scaling behavior of various hash maps for ticketing across different data distributions (see~\autoref{sec:setup} for workload details). The top row measures performance as throughput (higher is better) against thread count. The bottom row plots the speedup factor against thread count, measured as the single-threaded latency divided by the latency at a given thread count. Ideal scaling is the linear function $speedup=threads$, which is plotted as a dashed line. The vertical dashed line represents the threshold thread count between physical cores and SMT. Folklore* consistently achieves the best overall throughput across the tested range of thread counts and displays good scaling behavior.}
  \caption{Scaling behavior of various hash maps for ticketing. The top row measures performance as throughput (higher is better) against thread count. The bottom row plots the speedup factor against thread count, measured as the single-threaded latency divided by the latency at a given thread count. Ideal scaling is the linear function $speedup=threads$, which is plotted as a dashed line. The vertical dashed line represents the threshold thread count between physical cores and SMT.}
  \label{fig:scaling_ticketing}
\end{figure*}

\subsubsection*{\normalfont\bfseries Ticketing Performance} \label{sec:ticketing_evaluation}

In~\autoref{fig:scaling_ticketing}, we show the performance and scaling behavior of these designs across different cardinalities and data distributions (see~\autoref{sec:setup}). We find that Folklore* consistently performs best across all tested designs, despite its simple implementation. Crucially, Folklore* performance is resilient to data skew due to its fast path lookup. It also exhibits excellent scaling behavior for low cardinality workloads, although the speedup it achieves at high thread counts degrades as cardinality increases. At 48 threads (the number of real cores on the system), it achieves a 37.6x speedup for low cardinality, 22.8x for high cardinality and 13.8x with unique keys. SMT generally yields little benefit. 

Surprisingly, cuckoo hashing exhibits very poor scaling in high contention lookup workloads, despite us implementing a fast path and prior literature indicating good performance~\cite{li_algorithmic_2014}. We attribute this behavior to the use of fine-grained locks on buckets. Even though read locks are used on the fast path, bookkeeping by the locks may cause contention when multiple threads access the same resource in rapid succession. DashMap also uses read locks, which could explain its similar performance characteristics. 

\begin{figure}
  \centering
  \includegraphics[width=\linewidth]{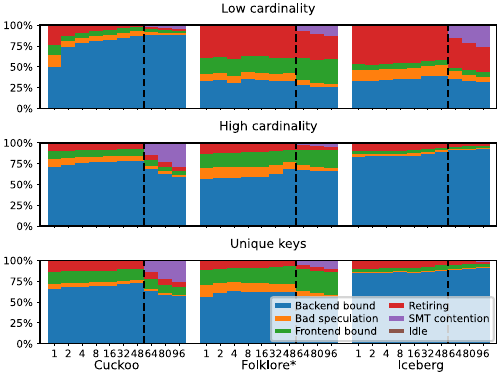}
  \Description{Breakdown of time spent on work by processor per workload and thread count. Shows both idle time and topdown analysis of time performing work. Shows the amount of time spent on work and the performance characteristics of the work. See~\autoref{sec:topdown} for more details on this mode of analysis.}
  \caption{Breakdown of time spent on work by processor per workload and thread count. Shows both idle time and topdown analysis of time performing work (see~\autoref{sec:topdown}).}
  \label{fig:topdown_map}
\end{figure}

In~\autoref{fig:topdown_map}, we perform topdown analysis~\cite{yasin_top-down_2014} on a selection of the hash table designs to understand the performance characteristics of each method. More detailed explanation of our methodology can be found in~\autoref{sec:topdown}. Our hypothesis regarding bookkeeping on locks aligns with the spike in backend bound behavior when multi-threading for cuckoo hashing on the low cardinality dataset (contention on atomics, used for bookkeeping in read locks, is considered backend bound). Similar behavior occurs for the heavy hitter dataset and for DashMap (not shown). 

In contrast, LeapMap and Folklore* are fully lock-free (and Iceberg is lock-free on its fast path lookup). These characteristics explain the gulf in scaling behavior in the low cardinality case, where speedup at high thread count is starkly higher for the latter three methods compared to the former two that use read locks. Further, by avoiding write locks with all methods, we avoid idle time from contention. \ul{Thus, it is essential to have a lockless ``fast path'' for reading previously-inserted values.}

\autoref{fig:topdown_map} also reveals that Folklore* shows less backend bound behavior compared to other methods, suggesting superior cache performance, a benefit of linear probing. In fact, when looking at raw performance counters, across the data distributions and thread counts (up to 48) in all but two cases Folklore* achieves the lowest number of cache misses, and in the majority has the lowest cache miss rate. For example, for the high cardinality dataset at 48 threads, Folklore* has 1.4e8 cache misses and a 39.9\% cache miss rate. The next best by total cache misses (Leapfrog) has considerably higher cache miss count and miss rate at 2.1e8 and 46.2\% respectively. That said, Folklore* still spends a substantial proportion of time backend bound at high cardinalities, explaining the poorer scaling as cardinality increases, as memory now becomes a bottleneck.

\subsubsection*{\normalfont\bfseries Discussion}
An interesting finding from the experiment results is that for our lookup and insert-only workload, even a simple linear probing design (Folklore*) achieves excellent performance. A contributing factor is that the typical downside to linear probing, deletions, is a non-issue given our workload. \ul{We conclude that to perform efficient ticketing, fancy hash tables are not required: linear probing is all you need.} 

One meta-explanation for Folklore*'s dominance is the benchmark workloads used by hash table researchers. Existing literature on concurrent hash table designs often do not test certain workloads important for our application. For example, neither of~\cite{li_algorithmic_2014, kelly_lock-free_2020} test a lookup-only workload (aligning with low cardinality workloads), and all workloads tested contained updates and deletions (of which there are none in our formulation of group aggregation). While this particular selection of workloads makes sense for a general-purpose hash table, the results do not apply to group aggregation. Put another way, the omission of read-heavy, delete-free workloads can cause the designs to be poorly optimized for our ticketing use case despite strong performance in mixed, general-purpose workloads. 

Even when benchmarking specifically for aggregation use cases, as in Maier et al.~\cite{maier_concurrent_2018}, chosen workloads often assume in-table aggregation (which is not possible without query compilation, since there are too many combinations of key types and aggregation functions), and is therefore update-heavy, not taking advantage of the lookup and insert-only semantics of ticketing. This misalignment has yielded conclusions that concurrent hash tables are still insufficient for aggregation purposes in the presence of skew, even when, in fact, they do not necessarily present a barrier. On the contrary, the simple nature of a lookup and insert workload can, for many existing concurrent hash table designs, make skewed workloads a non-issue. \ul{We recommend that database implementers use caution when evaluating benchmarks for general-purpose hash tables since group by aggregation has a distinct profile that is often overlooked.}

Finally, we make the observation that our definition of ticketing reduces the task to that of finding a perfect hash function; that is, the purpose of ticketing is to assign each key a unique, densely packed integer, which is exactly what a perfect hash function does. Gaffney and Patel~\cite{gaffney_is_2024} found that significant speedups can be achieved when integrating perfect hash functions into DuckDB's aggregation pipeline. Our formulation of ticketing aligns perfectly with this notion of perfect hashing, and ticketing could likely also greatly benefit from perfect hashing (i.e., ``skipping'' the ticketing step). Perfect hashing could essentially remove all contention from the ticketing phase, greatly accelerating fully concurrent aggregation.

\subsection{Partial Aggregate Update} \label{sec:update}
In the update step, we ``actually do'' the aggregation and modify the partial aggregate value corresponding to each ticket based on the associated row. Concurrency control problems that we avoided in the ticketing step have been deferred to this stage. In this section, we explore two classes of update methods, concurrent and thread local. We evaluate and discuss the methods and identify the situations where each exhibits good performance and scaling. 

\subsubsection*{\normalfont\bfseries Concurrent Update Method} A naive but general-purpose solution to managing concurrency in this step is to protect each cell of the vector of partial aggregate values with a lock. For each ticket, we acquire the lock on its cell in the partial aggregate vector, update the partial aggregate value, and release the lock. 

Another simple approach is to have each partial aggregate be an atomic. It is straightforward to perform aggregation functions such as \texttt{COUNT}, \texttt{SUM}, and \texttt{MIN/MAX}, but other more complicated aggregation functions may not be easy to implement with atomics. One general solution is to perform a lookup on the current partial aggregate, perform the update, and then use a \texttt{CAS} instruction to update the partial aggregate. However, this method increases the number of atomic operations, and hence contention, and also potentially suffers from the ABA problem~\cite{dechev_understanding_2010} (although this is probably unlikely for most aggregation functions). These fully concurrent update methods are simple to implement and very memory efficient, however, they suffer at higher levels of contention: if there is a heavy hitter in the data, every thread may simultaneously and repeatedly try to update the same partial aggregate value. There is potential for operation-specific optimizations, such as the inclusion of a fast-path for \texttt{MIN/MAX} that avoids an atomic write if not needed, but in the general case skewed workloads continue to present issues.

\subsubsection*{\normalfont\bfseries Thread Local Update Method}
To mitigate issues with contention, we explore a thread local approach to updates where each worker thread updates partial aggregates in its own thread local vector, eliminating all contention at the cost of a merge of all partial aggregate values at the end of the aggregation. Although the total work of the merge step scales with the number of threads, because the vectors of partial aggregates are all in the same order (ordered by ticket), the merge is trivially parallel and cache-efficient, which should in part mitigate the overhead.
Unfortunately, the work per thread does not decrease asymptotically as threads increase. Since each of $k$ threads is assigned $n/k$ tickets to merge (where $n$ is the number of unique keys), and there are $k$ threads worth of partial aggregates per ticket, each thread aggregates over $(n/k)*k$ partial aggregates, yielding $O(n)$ runtime.

Another challenge is that memory usage scales linearly with the number of threads and distinct keys, requiring a vector the size of the entire key space per thread. For very large datasets at high thread counts, this overhead could be a concern. One mitigating factor is the fact that vectors of partial aggregates are dense, as opposed to the hash tables allocated per thread in partitioned aggregation, reducing the gap between the two techniques' memory usage. We quantify the memory overhead in~\autoref{sec:memory}. 

\subsubsection*{\normalfont\bfseries Evaluation}
\begin{figure*}
  \centering
  \includegraphics[width=\linewidth]{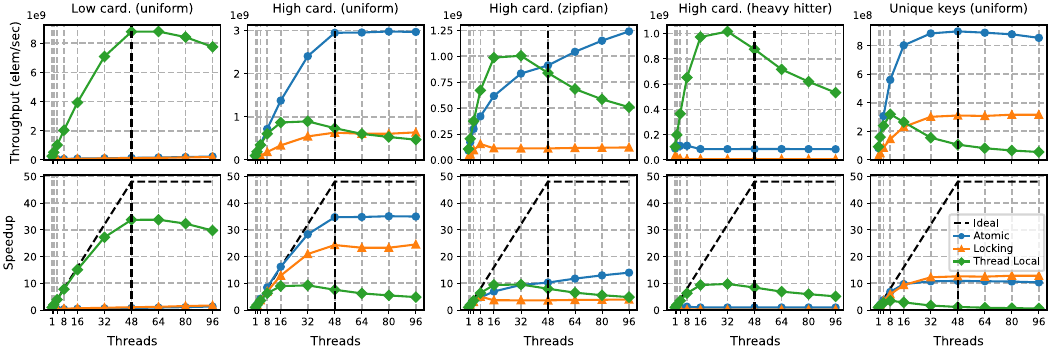}
  \Description{Scaling behavior of various partial aggregate update methods across different data distributions. Similarly to~\autoref{fig:scaling_ticketing}, the top row plots performance as throughput and the bottom row plots the speedup factor relative to single-threaded performance. Atomic updates display good performance and scaling in low contention scenarios while thread local updates have good performance and scaling for lower cardinality workloads. Thread local updates are also resilient to data skew.}
  \caption{Scaling behavior of various partial aggregate update methods across different data distributions.}
  \label{fig:scaling_update}
\end{figure*}

In~\autoref{fig:scaling_update}, we plot the performance of these update methods in isolation--that is, without ticketing. Keys are given as an integer from 0 to the max key, which is used directly as a ticket (i.e., a perfect hash function). We set up the experiment in this way to isolate the scaling behavior of each specific aggregation method. It is important to note that contention effects are far less pronounced than how they appear in these isolated benchmarks because the time it takes to ticket causes fewer threads to be at the update step at the same time, decreasing contention. 

Both thread local and atomic updates demonstrate strong scaling behavior in some workloads but neither is a clear winner across all workloads. Thread local updates have \ul{superior performance in the presence of high contention} (i.e., low cardinality datasets or skewed distributions) but \ul{degrade in performance as cardinality increases.} This is because, at lower numbers of unique keys, the underlying vectors of partial aggregates are small enough that there is minimal overhead from merging. At low cardinality, thread local updates achieve a substantial speedup of 33.7x at 48 threads. 

At higher cardinalities, the thread local method demonstrates inverse scaling behavior at higher thread counts, actually decreasing in throughput after 32 threads for the high cardinality case and 8 threads in unique key case. Still, at its peak, the thread local achieves a 9.2x speedup (at 32 threads) on the high cardinality case. The unique case has a peak speedup of only 3.5x (at 8 threads). Crucially, though, the \ul{speedup is consistent no matter the skew}--since the partial aggregates have no contention, skew can only help due to greater cache locality and leads to no performance degradation. This property makes thread local updates a good option even with relatively high cardinality workloads due to its consistency.

Meanwhile, \ul{atomic updates exhibit better scaling in workloads with low contention and high cardinality.} For the high cardinality and unique key workloads, when in the absence of skew, atomic updates are a clear winner. In fact, in the high cardinality case, atomics achieve a blazing 34.7x speedup at 48 threads. Surprisingly, atomic updates exhibit serviceable scaling even in the presence of some skew (achieving 10.3x speedup on the Zipfian workload) but fall short when the skew becomes too great (the heavy hitter workload). Fine-grained locked updates exhibits similar behavior to atomic updates (since it is bottlenecked by similar contention effects) but tends to be the worst performer due to its higher overhead. 

\subsubsection*{\normalfont\bfseries Discussion}
\begin{table}
  \caption{Desiderata fulfilled by each update method. For cardinality, skew, and thread count, the range of workloads where the given update method performs well in is given. For memory usage, the asymptotic behavior is given as a factor of $n$ (the number of unique keys) and $k$ (the number of threads).}
  \label{tab:update_desiderata}
  \begin{tabular}{c|ccc}
    \toprule
    & Atomic & Locking & Thread Local \\
    \midrule
    Cardinality & High-Unique & High-Unique & Low-High \\
    Skew & Low-Medium & Low & Low-High \\
    Threads & Low-High & Low-High & Low-Medium \\
    Memory & $O(n)$ & $O(n)$ & $O(kn)$ \\
    \bottomrule
\end{tabular}
\end{table}

Given the varied performance characteristics of these update methods, summarized in~\autoref{tab:update_desiderata}, ideal performance can be achieved by adaptively choosing the best method. Although imperfect, optimizers can estimate the number of unique keys~\cite{kipf_estimating_2019} to choose the update method on a per-query basis. 

There is a concerning performance gap for workloads that are both high cardinality and highly skewed. However, we note that the most extreme performance degradation occurs in rarer cases, such as when there are a particularly high number of duplicate values, or when elements are unique. Furthermore, it is important to note poor scaling in the partial aggregate update step does not preclude overall good performance from fully concurrent aggregation. In~\autoref{sec:stages_breakdown}, we find that the ticketing stage typically accounts for the bulk of overall execution time (since hash table operations tend to have much higher overhead than updates on a vector). Therefore, since we have found that ticketing exhibits good scaling behavior, provided that the poor scaling in the update step is not bad enough to be a bottleneck, the overall system still can scale well even with degraded performance in the update step. 

We believe that there is significant room for future work to improve this step. Particularly, a system that can combine both atomic or locked updates with thread local updates could take advantage of the benefits of both systems. This discussion bears similarity to hybrid aggregation approaches~\cite{cieslewicz_adaptive_2007, fent_practical_2021} that combine a shared hash table with thread local hash tables for heavy hitters. We also believe that there can be room for optimization by taking advantage of the vectorized nature of these updates. If locks can be obtained a vector at a time, we may be able to reduce the overhead from locking. 

\section{End-to-end experiments} \label{sec:experiments}
In this section, we combine the ticketing and update steps of fully concurrent aggregation and benchmark end-to-end performance. We also implement partitioned aggregation in the same environment and compare the results. For all fully concurrent aggregations, we use Folklore* for the ticketing step because it had the highest throughput across all tested cases. 

Recall that our experimental setup is described in~\autoref{sec:setup}. Our experimental analysis is broken into four sections. \textbf{Scaling (\ref{sec:scaling})} presents experiments measuring throughput and scaling behavior across different workloads. \textbf{Explaining Scaling (\ref{sec:behavior})} aims to explain the observed scaling behavior by analyzing time spent on each stage of aggregation and profiling results. \textbf{Other Platforms (\ref{sec:platforms})} replicates the main experiment on other architectures and discusses any differences in performance. \textbf{Tuple Size (\ref{sec:tuple_size})} varies the size of the aggregate tuple. \textbf{Resizing (\ref{sec:resizing})} compares aggregation methods in the case of poor cardinality estimation resulting in resizing behavior. \textbf{Memory Usage (\ref{sec:memory})} compares the peak memory usage of each aggregation method.

\subsection{Scaling} \label{sec:scaling}
\begin{figure*}
  \centering
  \includegraphics[width=\linewidth]{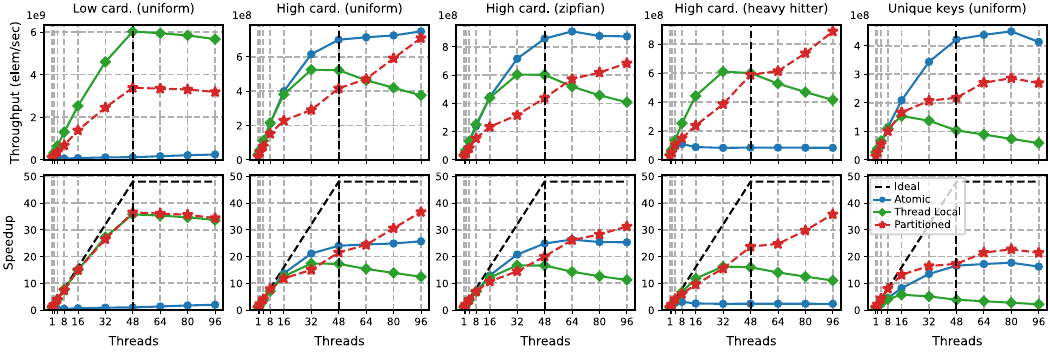}
  \Description{End-to-end evaluation of scaling behavior of fully concurrent aggregation methods vs. partitioned aggregation. Folklore* is used for the ticketing step for fully concurrent aggregation and both atomic and thread local updates are evaluated. Similar to~\autoref{fig:scaling_ticketing} and~\autoref{fig:scaling_update}, performance as throughput and the speedup factor relative to single-threaded execution are plotted against thread count.}
  \caption{End-to-end evaluation of scaling behavior of fully concurrent aggregation methods vs. partitioned aggregation. Folklore* is used for the ticketing step for fully concurrent aggregation.}
  \label{fig:scaling_e2e}
\end{figure*}

We evaluate scaling end-to-end (including both ticketing and partial aggregation update steps) in~\autoref{fig:scaling_e2e}. We test the atomic and thread local methods for fully concurrent aggregation and graph the results against those for partitioned aggregation using local preaggregation. The locked update approach is omitted because it performs strictly worse than atomics as a concurrent update method. 

\subsubsection*{\normalfont\bfseries Low Cardinality}
At low cardinality, fully concurrent thread local aggregation outperforms the other methods at all thread counts, achieving 1.78x throughput compared to the partitioned aggregation at 48 threads. As in~\autoref{sec:update}, atomic aggregation does not scale well. 

\subsubsection*{\normalfont\bfseries High Cardinality}
In the high cardinality case, we find that thread local aggregation has a clear advantage over partitioning at lower thread counts. At this cardinality, the local hash table used by partitioned aggregation spills most of its entries, which causes the partitioning to have to aggregate each value twice (once locally and once partition-wise). However, at high core counts the benefit narrows. For the non-heavy-hitter distribution, the atomic method also displays an advantage over partitioning, even in the presence of moderate skew in the Zipfian workload. 

Thread local aggregation does not stop scaling until reaching above 32 threads, despite doing so in the isolated update benchmarks from~\autoref{sec:update}. This behavior is because the update step is fast enough at low thread count such that the behavior of the ticketing step dominates the runtime, as seen in the performance breakdowns in~\autoref{sec:stages_breakdown}. 

Surprisingly, partitioned aggregation continues to scale when thread count exceeds the number of physical cores. However, even at the high range of thread counts, partitioned aggregation is around parity or slightly worse than fully concurrent methods. The exception is the heavy hitter case, where partitioned aggregation becomes a clear winner as a result of the performance gains from local preaggregation (which captures all the heavy hitter entries). However, this is a relatively extreme case of preaggregation effects.

\subsubsection*{\normalfont\bfseries Unique Key}
In the pure insert workload, fully concurrent aggregation also exhibits superior performance at high thread count, but only when using atomic updates. Thread local aggregation scales reasonably well at lower thread counts, but performance significantly degrades as the number of threads increases. Therefore, implementations using thread local updates should note to limit the maximum thread count under certain workload conditions. In this dataset, since there is no contention on the underlying partial aggregations, using atomics creates very little overhead. 

\subsubsection*{\normalfont\bfseries Comparisons}
Notably, the performance advantage of fully concurrent aggregation against partitioned aggregation does not come from its scaling behavior, with fully concurrent aggregation achieving excellent scaling at high thread counts. However, fully concurrent aggregation has significantly lower hash table lookup overhead, since it only inserts into the ticketing hash table a single time (rather than performing aggregation locally and partition-wise).  

In almost all cases, a form of fully concurrent aggregation achieves parity or better, with the only exception being high cardinality with heavy hitters at high thread counts due to the poor performance of atomic aggregation in the presence of skew. However, our heavy hitter case is a particularly extreme case of skew with 50\% the same value. More realistic Zipfian-distributed workloads can be handled by atomic updates without too much issue. 

Furthermore, the fact that thread local aggregation achieves its maximum speedup at a relatively low number of threads is advantageous in many real-world applications. It is important in practice to actually impose a max thread count for the thread local case, freeing up resources for other queries. \ul{Practitioners should consider if using twice as many resources (on partitioning) for a small lift in throughput is worthwhile,} or if those additional resources are better allocated to more concurrent queries. We imagine that this tradeoff is acceptable for some, but not all, systems.

\subsubsection*{\normalfont\bfseries Recommendations}
As we discussed in~\autoref{sec:update}, \ul{ideally the method for partial aggregate updates should be chosen based on workload.} Given that in almost all cases, some form of fully concurrent aggregation achieves markedly superior performance over partitioning, a dynamic system could yield significant speedup over current practice. However, perhaps one of the more important findings of this paper is that \ul{no single aggregation method (including partitioning) is desirable in all cases.} That being said, thread local aggregation shows significant versatility. It has acceptable tail behavior and superior throughput in many workloads. Furthermore, its excellent performance at low thread counts across all workloads makes it ideal for cases where resources are shared or limited. \ul{Therefore, if choosing only one aggregation method for a system, fully concurrent aggregation with thread local updates is a strong candidate.} While we do not endorse one single aggregation method as the definitive choice for a database, this work on fully concurrent aggregation expands the toolkit available for designing a well-optimized query executor. 

\subsection{Explaining Behavior} \label{sec:behavior}
In this section, we attempt to explain the scaling behavior of each aggregation method. We first take a high-level look by breaking down the time spent on each task (ticketing, partial aggregate update, etc). Then, we analyze specific performance counters for a deeper understanding of how execution characteristics change with thread count. 

\subsubsection*{\normalfont\bfseries Task Breakdown}  \label{sec:stages_breakdown} 
\begin{figure}
  \centering
  \includegraphics[width=\linewidth]{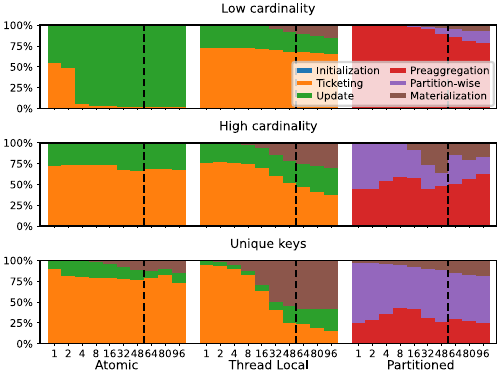}
  \Description{Percent of time spent on each step of aggregation. Ticketing consistently takes much more time than the update step except in high contention workloads atomic update workloads, which becomes a bottleneck. Materialization becomes a significant part of thread local aggregation’s runtime at high cardinality and thread count.}
  \caption{Percent of time spent on each step of aggregation.}
  \label{fig:stages_breakdown}
\end{figure}

First, we time each stage of the aggregation process and plot the proportion of time spent in~\autoref{fig:stages_breakdown}. The ticketing and update stages are as discussed in~\autoref{sec:fca} and the preaggregation and partition-wise aggregation stages are as discussed in~\autoref{sec:part_agg}. The materialization stage consists of the work to turn results into a columnar format that can be pushed to the next query operator, including the cost of merging thread local partial aggregates and combining each partition's aggregated values. 

We find that for fully concurrent aggregations, ticketing typically takes significantly more time than updating the partial aggregates. The exception is the low cardinality case for atomic updates, which at 4 threads becomes the major performance bottleneck. Therefore, the choice of update step method should be driven more by tail performance. As long as it does not become a performance bottleneck, \ul{even relatively poor scaling in the update step is not insurmountable.} This is the primary factor that allows thread local updates to maintain relatively good performance. 

Further, we find that the thread local method's materialization cost becomes increasingly significant as thread count increases, which aligns with our experimental results and theoretical findings from~\autoref{sec:update}. Materialization is fast enough in many cases that its poor scaling is not a bottleneck. However, at high thread counts with very high cardinalities, the materialization step can become the driving factor for degraded performance.

The materialization effects of partitioned aggregation are also particularly interesting. The materialization of each partition-wise hash table and copying of values to the final vector of all outputs is parallelized. The decrease in materialization time at very high thread counts appears to contribute to the improved scaling of partitioned aggregation even after thread count exceeds real core count. We attribute some of this odd behavior to the size of the hash table of partition-wise aggregates, which must be walked in order to materialize the key and value output columns. The size of each partition's hash table decreases with the number of partitions.

\subsubsection*{\normalfont\bfseries Performance Counters} \label{sec:topdown}
\begin{figure}
  \centering
  \includegraphics[width=\linewidth]{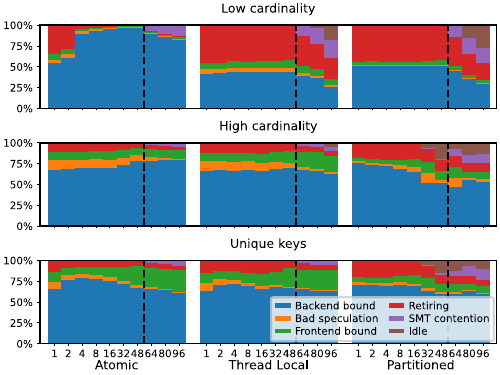}
  \Description{Breakdown of time spent on work by processor per workload and thread count. Shows both idle time and topdown analysis of time performing work.}
  \caption{Breakdown of time spent on work by processor per workload and thread count. Shows both idle time and topdown analysis of time performing work.}
  \label{fig:topdown_e2e}
\end{figure}

\begin{figure}
  \includegraphics[width=\linewidth]{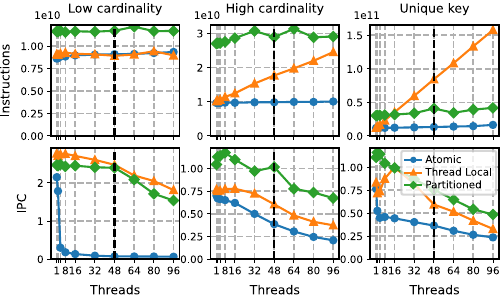}
  \Description{Instruction count and IPC vs. threads. Contributes to understanding the performance characteristics of each workload and the amount of work to be performed.}
  \caption{Instruction count and IPC vs. threads.}
  \label{fig:instructions_e2e}
\end{figure}

We also gather performance counter data for each workload using Linux's \texttt{perf} tool to observe how execution characteristics vary with thread count. Analysis of the resulting data is made difficult by the complicated nature of the workload, whose performance is driven by multiple factors, including contention, time complexity scaling, etc. We therefore take a layered approach and separate the analysis into three questions:

\begin{enumerate}
\item \textbf{Q1: How much time is spent performing ``work.''} Specifically, we measure the proportion of time a worker thread is scheduled on a core. The compliment of this time is idle time, which occurs due to lock contention, synchronization effects, or lack of resources. 
\item\textbf{Q2: Of the time spent on ``work,'' what are the performance characteristics?} These performance characteristics impact the rate that the processor performs work, that is instructions per cycle (IPC). We also employ topdown analysis~\cite{yasin_top-down_2014} for granular insights into the bottlenecks. 
\item\textbf{Q3: How much ``work'' is there?} We measure the amount of work with total instruction count. 
\end{enumerate}

In~\autoref{fig:topdown_e2e} we break down work time with topdown analysis as well as plot the split of time working and idling (Q1, Q2). In~\autoref{fig:instructions_e2e} we plot the instructions and IPC (Q2, Q3).

At a low cardinality, we see that atomic updates quickly becomes backend bound (contention from atomic loads is considered a backend bound effect, not idle time, unlike lock contention). In contrast, the thread local and partitioned methods show relative stability in its execution characteristics until idle time and SMT contention start to become a factor. In addition, a high proportion of instructions are retiring, indicating that execution is compute-bound and not bottlenecked by memory, speculation effects, etc.

At higher cardinalities, more work is spent backend bound across the board. This behavior is expected given that at high cardinalities, every hash table lookup and partial aggregate update is expected to be a cache miss. We also see some idle time occurring at higher thread counts for the partitioned method, even prior to SMT. This observation points to a theoretical drawback of the approach. Partitioning tends to be more prone to synchronization issues because all threads must wait for the final partition to complete its task. Meanwhile, work-stealing is generally difficult in such a setup.

Now turning to the instruction and IPC counts in~\autoref{fig:instructions_e2e}, we confirm that the total work performed by thread local updates increases with thread count due to materialization. Also revealing is that while the base instruction count of partitioned aggregation stays stable, it is considerably higher than the other methods, owing to partitioning's added overhead. The counterbalancing effect is its relatively higher IPC at higher cardinalities. This behavior helps to explain why SMT benefits partitioned aggregation in particular. The high instruction count and IPC indicates that the processor is better able to utilize the time spent stalling on memory or other resources to continue execution on other compute-bound work.

\subsection{Other Platforms} \label{sec:platforms}
\begin{figure}
  \centering
  \includegraphics[width=\linewidth]{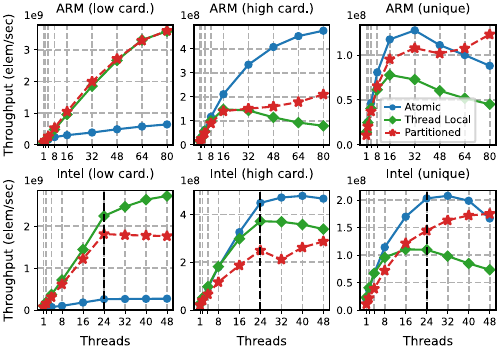}
  \Description{Scaling behavior of aggregation on ARM (Ampere Altra Q80-30) and Intel (Xeon Gold 5412U) machines. Extended benchmarks and profiling results on skewed datasets can be found at~\href{https://rm.cab/ghplots}{rm.cab/ghplots}.}
  \caption{Scaling behavior of aggregation on ARM (Ampere Altra Q80-30) and Intel (Xeon Gold 5412U) machines. Extended benchmarks and profiling results on skewed datasets can be found at~\href{https://rm.cab/ghplots}{rm.cab/ghplots}.}
  \label{fig:platforms}
\end{figure}

In addition to the AMD platform, we also ran experiments on machines with ARM and Intel processors. The different hardware characteristics of each are described in~\autoref{sec:setup}. We graph the throughput achieved on these systems in~\autoref{fig:platforms}. We find that although the best technique for a given configuration may vary, the overall shape of the graph remains similar to our results in~\autoref{fig:scaling_e2e}. 

On ARM, we find that partitioned aggregations performs particularly poorly on the high cardinality case. We found that on this system, a larger amount of time was spent on materialization, which combines all the thread local hash tables into two output columns of keys and values. This step is almost purely reads and writes to memory. Therefore, we believe this poor performance could relate to the lower memory bandwidth of our ARM setup, which is 3200MT/s as compared to 4800MT/s on our AMD setup and 5600MT/s on our Intel setup.

Note that the Intel chip has fewer cores, so the graphs do not reveal severe performance degradation for thread local aggregation at high thread counts. Here, the performance of thread local updates is excellent in all but the unique case, and even then performs admirably. Atomic updates, in turn, perform much better than partitioned at higher cardinalities. These results reinforce our discussion that on more resource-constrained machines, fully concurrent aggregation is an especially convincing choice.

\subsection{Tuple Size} \label{sec:tuple_size}
\begin{figure}
  \centering
  \includegraphics[width=\linewidth]{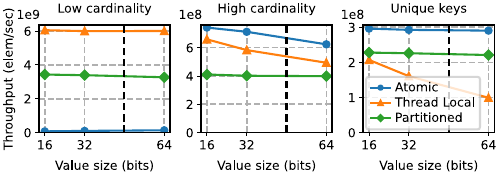}
  \Description{Throughput graphed against size in bits of value. For higher cardinality workloads, we see increasing the size of the aggregated value decreases throughput significantly for thread local aggregation, but marginally for the rest.}
  \caption{Throughput vs. size of value being aggregated.}
  \label{fig:tuple_size}
\end{figure}

Up until now, we assumed that all values were 64-bit integers. We now investigate the performance impact of the size of the tuple. Specifically, we hold the thread count (48) and key size (64-bit) constant and vary the value size to 16-bit, 32-bit, and 64-bit integers. We choose vary the output size because it impacts the update step, whereas the key size mostly influences the speed of hashing. We plot throughput against value size in~\autoref{fig:tuple_size}.

We find that the effect on performance is most pronounced for the higher cardinality workloads, which aligns with their backend bound behavior and thus sensitivity to factors such as memory bandwidth. The most significant impact is on thread local aggregation, which has clearly decreasing throughput at higher value sizes, especially in the unique keys case. This is because materialization of thread local aggregates requires many sequential reads from memory. The total number of reads during this operation increases with tuple size. On the other hand, atomic updates and partitioned aggregation have mostly random memory accesses, which incur a cache miss regardless of tuple size.

\subsection{Resizing} \label{sec:resizing}
A particularly thorny challenge with concurrent hash tables is how to resize them efficiently. In most cases, resizing requires all other threads to pause work to accommodate reallocation and migration. Although cardinality estimation can often allow a properly sized initial allocation~\cite{kipf_estimating_2019}, poor estimates are always a possibility.

To test the impact of hash table resizing, we adopt Maier et al.'s~\cite{maier_concurrent_2018} method for contention-less fully concurrent migration of hash table entries for Folklore*. In this experiment, we set the capacity of the ticketing hash table and partial aggregate vectors to be half of the required capacity, forcing a resize.  

\begin{figure}[h] 
  \centering
  \includegraphics[width=\linewidth]{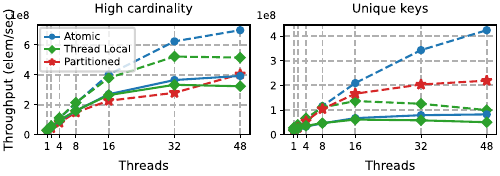}
  \Description{Scaling behavior of fully concurrent aggregation with a resize. Dashed lines indicate throughput without resizing and solid lines with. Fully concurrent aggregation exhibits far more performance degradation under these circumstances. Partitioned approach is not dependent on accurate cardinality estimation, but it is plotted for comparison.}
  \caption{Scaling behavior of fully concurrent aggregation with resize. Dashed lines are without resizing and solid lines are with. Partitioned plotted for reference.}
  \label{fig:resizing}
\end{figure}

\autoref{fig:resizing} shows that the fully concurrent workload is does display significant performance degradation at higher thread counts in the presence of resizing. Resizing causes a considerable but not insurmountable latency increase for the high cardinality case. At 48 threads, atomics increase in latency by 1.8x and thread local by 1.6x. However, in the very worst case, unique keys, the degradation is much more extreme with a 5.2x latency increase for atomics and 2.0x for thread local. Thus, while not a show stopper, \ul{fully concurrent aggregation appears to be sensitive to the cost of resizing,} especially in the most extreme cardinality cases. Future work on improving the performance of such resizes has significant headroom for improvement. Resizing performance and should be an important dimension for analyzing the suitability of concurrent hash table designs for database applications, and we hope this work motivates more research in the area. 

\subsection{Memory} \label{sec:memory}
\begin{table*}
  \caption{Peak memory usage of different aggregation methods, measured in GB.}
  \label{tab:memory}
  \begin{tabular}{c|ccc|ccc|ccc}
    \toprule
    Threads & \multicolumn{3}{c|}{1} & \multicolumn{3}{c|}{8} & \multicolumn{3}{c}{48} \\
    Method & Atomic & Thread Local & Partitioned & Atomic & Thread Local & Partitioned & Atomic & Thread Local & Partitioned \\
    \midrule
    Low card. & 0.001 & 0.001 & 0.000 & 0.004 & 0.004 & 0.004 & 0.024 & 0.028 & 0.024 \\
    High card. & 0.299 & 0.299 & 2.898 & 0.302 & 0.824 & 3.133 & 0.322 & 3.828 & 2.971 \\
    Unique keys & 4.098 & 4.098 & 4.625 & 4.102 & 9.317 & 5.000 & 4.121 & 39.143 & 5.137 \\
    \bottomrule
  \end{tabular}
\end{table*}

Our model of execution assumes that all data structures fit within memory. It is not obvious how to adapt fully concurrent aggregation to disk spilling, unlike partitioning-based approaches~\cite{kuiper_robust_2024}. Therefore, understanding the severity of memory usage can be relevant for the feasibility of aggregation over large datasets. 

In~\autoref{sec:update}, we note that thread local updates create significant memory overhead while atomic updates are very space efficient. Analyzing the partitioned method is less clear, since much depends on the spilling behavior. However, in the very worst case, when almost all keys are spilled, memory usage is bounded by the total number of elements (not just unique keys), which can severely increase memory usage.

Comparing the peak memory usage of each technique in~\autoref{tab:memory}, we find that our theoretical model holds. Atomic aggregation performs best, while partitioned aggregation displays very high memory overhead for higher cardinality cases. Thread local aggregation exhibits surprisingly good memory usage characteristics, using much fewer memory resources than partitioned aggregation until the high end of thread count. In the unique case, at very high thread counts, thread local aggregation has much higher memory usage. However, since we would limit the number of threads in practice, the memory usage would not reach the extreme case. Although memory considerations should be carefully considered on a system and workload basis, our results indicate that \ul{fully concurrent aggregation has substantial advantages over partitioning across many cases, although it is not obvious how to spill to disk.}

\section{Related Work} \label{sec:related}

The performance differences between concurrent aggregation using a shared hash table versus a partitioned approach using a local aggregation table have been benchmarked and studied extensively over the years~\cite{cieslewicz_adaptive_2007, ye_scalable_2011, luan_experimental_2022}. Notably, prior work does not propose indirection to reduce hash table operations to a lookup and insert-only workload, instead performing updates within the hash table itself, which requires significant concurrency control. As a result, the consensus in the literature is that thread local aggregation is not feasible due to contention costs in the presence of skew. One exception is~\cite{romanous_efficient_2021}, which used indirection, but in the context of FPGAs. Unlike our work,~\cite{romanous_efficient_2021} performs updates in the hash table but prevents simultaneous access using a hardware cache (not adaptable to general-purpose CPUs) to synchronize all update requests on the same key. 

Research from outside the database domain on concurrent hash table designs are also relevant to our work. In~\autoref{sec:ticketing}, we tested state-of-the-art hash table designs from or related to those described in prior work~\cite{li_algorithmic_2014, pandey_iceberght_2023, maier_concurrent_2018, kelly_lock-free_2020}. However, as previously noted, even prior work that consider hash aggregation did not customize their tables to only support the required operations, instead focusing on general-purpose workloads. While the body of work on concurrent hash tables is informative, each design must be critically reevaluated in the context of our specialized use case to make a determination about their performance for aggregations. 

In addition to the use of shared hash tables, many other methods for aggregation have been proposed. The method we compare most directly to is partitioned methods, which has been successfully integrated into many real-world systems~\cite{raman_db2_2013, leis_morsel-driven_2014, kuiper_robust_2024, lamb_aggregating_2023}. Various hybrid approaches that leverage local aggregations for heavy hitters and a global shared table for other values have been proposed in order to balance performance and memory concerns~\cite{cieslewicz_adaptive_2007, fent_practical_2021}. Ideas from these works may be beneficial for resolving the challenges we found with our fully concurrent aggregation at high thread counts and cardinalities. Sort-based aggregations, which are exceptionally efficient when data is already sorted~\cite{lamb_vertica_2012}, have also been extensively evaluated~\cite{vaghasiya_hybrid_2024, vaghasiya_experiments_2024, graefe_sort_1994}. Müller et al.~\cite{muller_cache-efficient_2015} note that when aggregating keys larger than CPU registers, sorting and hashing are computationally similar. In the Google F1 system, where distinct value counts are especially difficult to estimate, Do et al.~\cite{do_efficient_2022} showed that specialized sorting techniques could be more robust than hash-based aggregation. Sort-based aggregations also have operational advantages in out-of-core databases, as an external sort is generally simpler than external hashing~\cite{vaghasiya_hybrid_2024,do_efficient_2022}.

While this work focused on morsel-driven systems, other execution models, such as operator-level parallelism~\cite{graefe_volcano_1993, lamb_aggregating_2023} and pipeline parallelism~\cite{zaharia_spark_2010}, are also popular. It seems plausible that different aggregation methods may be optimal under different execution models, and we leave such investigations to future work.

Finally, a closely-related operation to hash aggregations is hash joins. Similar debates about partitioning have occurred in the literature~\cite{bandle_partition_2021}, and specialized hash tables (multi-maps) tend to have large performance benefits~\cite{leis_morsel-driven_2014, birler_simple_2024}. This body of work has had a similar overarching takeaway that specialized hash tables are crucial to achieving good performance, but their methods are generally incompatible with the needs of fully concurrent aggregation (e.g., multi-maps, probing). 

\section{Conclusions and future work} \label{sec:conclusion}
The go-to group by aggregation technique in modern analytic DBMSes is partitioning-based. In this work, we make the case for a simpler approach using a global concurrent hash table. For this simpler approach to match the performance of purpose-built partitioning techniques, a purpose-built hash table is required. We explored the operational benefits and drawbacks of both approaches and offered guidance to implementers along the way.

In future work, we plan to investigate more complex hash table designs as well as hybrid aggregation strategies (e.g., with teams of threads sharing an aggregation table to reduce contention). Another promising direction for investigation is understanding how new hardware features--like vectorized atomic operations--could be used to accelerate group aggregation even further. Additionally, the fully concurrent approach we analyze here suffers from issues with resizing and spilling, which are ripe areas for future work. Finally, this work only examined hash aggregation in a morsel-driven execution model: investigations into sort-based techniques, along with other execution models, could also yield interesting results.

\clearpage

\bibliographystyle{ACM-Reference-Format}
\bibliography{references}


\begin{thebibliography}{36}


\ifx \showCODEN    \undefined \def \showCODEN     #1{\unskip}     \fi
\ifx \showDOI      \undefined \def \showDOI       #1{#1}\fi
\ifx \showISBNx    \undefined \def \showISBNx     #1{\unskip}     \fi
\ifx \showISBNxiii \undefined \def \showISBNxiii  #1{\unskip}     \fi
\ifx \showISSN     \undefined \def \showISSN      #1{\unskip}     \fi
\ifx \showLCCN     \undefined \def \showLCCN      #1{\unskip}     \fi
\ifx \shownote     \undefined \def \shownote      #1{#1}          \fi
\ifx \showarticletitle \undefined \def \showarticletitle #1{#1}   \fi
\ifx \showURL      \undefined \def \showURL       {\relax}        \fi
\providecommand\bibfield[2]{#2}
\providecommand\bibinfo[2]{#2}
\providecommand\natexlab[1]{#1}
\providecommand\showeprint[2][]{arXiv:#2}

\bibitem[\protect\citeauthoryear{Bandle, Giceva, and Neumann}{Bandle et~al\mbox{.}}{2021}]%
        {bandle_partition_2021}
\bibfield{author}{\bibinfo{person}{Maximilian Bandle}, \bibinfo{person}{Jana Giceva}, {and} \bibinfo{person}{Thomas Neumann}.} \bibinfo{year}{2021}\natexlab{}.
\newblock \showarticletitle{To {Partition}, or {Not} to {Partition}, {That} is the {Join} {Question} in a {Real} {System}}. In \bibinfo{booktitle}{\emph{Proceedings of the 2021 {International} {Conference} on {Management} of {Data}}}. \bibinfo{publisher}{ACM}, \bibinfo{address}{Virtual Event China}, \bibinfo{pages}{168--180}.
\newblock
\urldef\tempurl%
\url{https://doi.org/10.1145/3448016.3452831}
\showDOI{\tempurl}


\bibitem[\protect\citeauthoryear{Bender, Conway, Farach-Colton, Kuszmaul, and Tagliavini}{Bender et~al\mbox{.}}{2023}]%
        {bender_iceberg_2023}
\bibfield{author}{\bibinfo{person}{Michael~A. Bender}, \bibinfo{person}{Alex Conway}, \bibinfo{person}{Martín Farach-Colton}, \bibinfo{person}{William Kuszmaul}, {and} \bibinfo{person}{Guido Tagliavini}.} \bibinfo{year}{2023}\natexlab{}.
\newblock \showarticletitle{Iceberg {Hashing}: {Optimizing} {Many} {Hash}-{Table} {Criteria} at {Once}}.
\newblock \bibinfo{journal}{\emph{J. ACM}} \bibinfo{volume}{70}, \bibinfo{number}{6} (\bibinfo{date}{Dec.} \bibinfo{year}{2023}), \bibinfo{pages}{1--51}.
\newblock
\showISSN{0004-5411, 1557-735X}
\urldef\tempurl%
\url{https://doi.org/10.1145/3625817}
\showDOI{\tempurl}


\bibitem[\protect\citeauthoryear{Birler, Schmidt, Fent, and Neumann}{Birler et~al\mbox{.}}{2024}]%
        {birler_simple_2024}
\bibfield{author}{\bibinfo{person}{Altan Birler}, \bibinfo{person}{Tobias Schmidt}, \bibinfo{person}{Philipp Fent}, {and} \bibinfo{person}{Thomas Neumann}.} \bibinfo{year}{2024}\natexlab{}.
\newblock \showarticletitle{Simple, {Efficient}, and {Robust} {Hash} {Tables} for {Join} {Processing}}. In \bibinfo{booktitle}{\emph{Proceedings of the 20th {International} {Workshop} on {Data} {Management} on {New} {Hardware}}}. \bibinfo{publisher}{ACM}, \bibinfo{address}{Santiago AA Chile}, \bibinfo{pages}{1--9}.
\newblock
\urldef\tempurl%
\url{https://doi.org/10.1145/3662010.3663442}
\showDOI{\tempurl}


\bibitem[\protect\citeauthoryear{Boncz, Zukowski, and Nes}{Boncz et~al\mbox{.}}{2005}]%
        {boncz_monetdbx100_2005}
\bibfield{author}{\bibinfo{person}{Peter~A. Boncz}, \bibinfo{person}{Marcin Zukowski}, {and} \bibinfo{person}{Niels Nes}.} \bibinfo{year}{2005}\natexlab{}.
\newblock \showarticletitle{{MonetDB}/{X100}: {Hyper}-{Pipelining} {Query} {Execution}}. \bibinfo{publisher}{www.cidrdb.org}, \bibinfo{address}{Asilomar, CA, USAW}, \bibinfo{pages}{225--237}.
\newblock
\urldef\tempurl%
\url{http://cidrdb.org/cidr2005/papers/P19.pdf}
\showURL{%
\tempurl}


\bibitem[\protect\citeauthoryear{Cieslewicz and Ross}{Cieslewicz and Ross}{2007}]%
        {cieslewicz_adaptive_2007}
\bibfield{author}{\bibinfo{person}{John Cieslewicz} {and} \bibinfo{person}{Kenneth~A. Ross}.} \bibinfo{year}{2007}\natexlab{}.
\newblock \showarticletitle{Adaptive aggregation on chip multiprocessors}. In \bibinfo{booktitle}{\emph{Proceedings of the 33rd {International} {Conference} on {Very} {Large} {Data} {Bases}}} \emph{(\bibinfo{series}{{VLDB} '07})}. \bibinfo{publisher}{VLDB Endowment}, \bibinfo{address}{Vienna, Austria}, \bibinfo{pages}{339--350}.
\newblock
\showISBNx{978-1-59593-649-3}


\bibitem[\protect\citeauthoryear{Clucas}{Clucas}{[n.d.]}]%
        {clucas_leapfrog_nodate}
\bibfield{author}{\bibinfo{person}{Rob Clucas}.} \bibinfo{year}{[n.d.]}\natexlab{}.
\newblock \bibinfo{title}{Leapfrog}.
\newblock
\newblock
\urldef\tempurl%
\url{https://github.com/robclu/leapfrog}
\showURL{%
\tempurl}


\bibitem[\protect\citeauthoryear{Dechev, Pirkelbauer, and Stroustrup}{Dechev et~al\mbox{.}}{2010}]%
        {dechev_understanding_2010}
\bibfield{author}{\bibinfo{person}{Damian Dechev}, \bibinfo{person}{Peter Pirkelbauer}, {and} \bibinfo{person}{Bjarne Stroustrup}.} \bibinfo{year}{2010}\natexlab{}.
\newblock \showarticletitle{Understanding and {Effectively} {Preventing} the {ABA} {Problem} in {Descriptor}-{Based} {Lock}-{Free} {Designs}}. In \bibinfo{booktitle}{\emph{2010 13th {IEEE} {International} {Symposium} on {Object}/{Component}/{Service}-{Oriented} {Real}-{Time} {Distributed} {Computing}}}. \bibinfo{publisher}{IEEE}, \bibinfo{address}{Carmona, Spain}, \bibinfo{pages}{185--192}.
\newblock
\urldef\tempurl%
\url{https://doi.org/10.1109/isorc.2010.10}
\showDOI{\tempurl}


\bibitem[\protect\citeauthoryear{Do, Graefe, and Naughton}{Do et~al\mbox{.}}{2022}]%
        {do_efficient_2022}
\bibfield{author}{\bibinfo{person}{Thanh Do}, \bibinfo{person}{Goetz Graefe}, {and} \bibinfo{person}{Jeffrey Naughton}.} \bibinfo{year}{2022}\natexlab{}.
\newblock \showarticletitle{Efficient {Sorting}, {Duplicate} {Removal}, {Grouping}, and {Aggregation}}.
\newblock \bibinfo{journal}{\emph{ACM Transactions on Database Systems}} \bibinfo{volume}{47}, \bibinfo{number}{4} (\bibinfo{date}{Dec.} \bibinfo{year}{2022}), \bibinfo{pages}{1--35}.
\newblock
\showISSN{0362-5915, 1557-4644}
\urldef\tempurl%
\url{https://doi.org/10.1145/3568027}
\showDOI{\tempurl}


\bibitem[\protect\citeauthoryear{Durner, Leis, and Neumann}{Durner et~al\mbox{.}}{2019}]%
        {durner_impact_2019}
\bibfield{author}{\bibinfo{person}{Dominik Durner}, \bibinfo{person}{Viktor Leis}, {and} \bibinfo{person}{Thomas Neumann}.} \bibinfo{year}{2019}\natexlab{}.
\newblock \showarticletitle{On the {Impact} of {Memory} {Allocation} on {High}-{Performance} {Query} {Processing}}. In \bibinfo{booktitle}{\emph{Proceedings of the 15th {International} {Workshop} on {Data} {Management} on {New} {Hardware}}}. \bibinfo{publisher}{ACM}, \bibinfo{address}{Amsterdam Netherlands}, \bibinfo{pages}{1--3}.
\newblock
\urldef\tempurl%
\url{https://doi.org/10.1145/3329785.3329918}
\showDOI{\tempurl}


\bibitem[\protect\citeauthoryear{Fent and Neumann}{Fent and Neumann}{2021}]%
        {fent_practical_2021}
\bibfield{author}{\bibinfo{person}{Philipp Fent} {and} \bibinfo{person}{Thomas Neumann}.} \bibinfo{year}{2021}\natexlab{}.
\newblock \showarticletitle{A practical approach to groupjoin and nested aggregates}.
\newblock \bibinfo{journal}{\emph{Proceedings of the VLDB Endowment}} \bibinfo{volume}{14}, \bibinfo{number}{11} (\bibinfo{date}{July} \bibinfo{year}{2021}), \bibinfo{pages}{2383--2396}.
\newblock
\showISSN{2150-8097}
\urldef\tempurl%
\url{https://doi.org/10.14778/3476249.3476288}
\showDOI{\tempurl}
\newblock
\shownote{Publisher: Association for Computing Machinery (ACM).}


\bibitem[\protect\citeauthoryear{Gaffney and Patel}{Gaffney and Patel}{2024}]%
        {gaffney_is_2024}
\bibfield{author}{\bibinfo{person}{Kevin~P. Gaffney} {and} \bibinfo{person}{Jignesh~M. Patel}.} \bibinfo{year}{2024}\natexlab{}.
\newblock \showarticletitle{Is {Perfect} {Hashing} {Practical} for {OLAP} {Systems}?} \bibinfo{publisher}{www.cidrdb.org}, \bibinfo{address}{Chaminade, HI, USA}.
\newblock
\urldef\tempurl%
\url{https://www.cidrdb.org/cidr2024/papers/p65-gaffney.pdf}
\showURL{%
\tempurl}


\bibitem[\protect\citeauthoryear{Graefe, Linville, and Shapiro}{Graefe et~al\mbox{.}}{1994}]%
        {graefe_sort_1994}
\bibfield{author}{\bibinfo{person}{G. Graefe}, \bibinfo{person}{A. Linville}, {and} \bibinfo{person}{L.D. Shapiro}.} \bibinfo{year}{1994}\natexlab{}.
\newblock \showarticletitle{Sort vs. hash revisited}.
\newblock \bibinfo{journal}{\emph{IEEE Transactions on Knowledge and Data Engineering}} \bibinfo{volume}{6}, \bibinfo{number}{6} (\bibinfo{date}{Dec.} \bibinfo{year}{1994}), \bibinfo{pages}{934--944}.
\newblock
\showISSN{10414347}
\urldef\tempurl%
\url{https://doi.org/10.1109/69.334883}
\showDOI{\tempurl}


\bibitem[\protect\citeauthoryear{Graefe and McKenna}{Graefe and McKenna}{1993}]%
        {graefe_volcano_1993}
\bibfield{author}{\bibinfo{person}{G. Graefe} {and} \bibinfo{person}{W.J. McKenna}.} \bibinfo{year}{1993}\natexlab{}.
\newblock \showarticletitle{The {Volcano} optimizer generator: extensibility and efficient search}. In \bibinfo{booktitle}{\emph{Proceedings of {IEEE} 9th {International} {Conference} on {Data} {Engineering}}}. \bibinfo{publisher}{IEEE Comput. Soc. Press}, \bibinfo{address}{Vienna, Austria}, \bibinfo{pages}{209--218}.
\newblock
\urldef\tempurl%
\url{https://doi.org/10.1109/icde.1993.344061}
\showDOI{\tempurl}


\bibitem[\protect\citeauthoryear{Kelly, Pearlmutter, and Maguire}{Kelly et~al\mbox{.}}{2020}]%
        {kelly_lock-free_2020}
\bibfield{author}{\bibinfo{person}{Robert Kelly}, \bibinfo{person}{Barak~A. Pearlmutter}, {and} \bibinfo{person}{Phil Maguire}.} \bibinfo{year}{2020}\natexlab{}.
\newblock \showarticletitle{Lock-{Free} {Hopscotch} {Hashing}}.
\newblock In \bibinfo{booktitle}{\emph{Symposium on {Algorithmic} {Principles} of {Computer} {Systems}}}. \bibinfo{publisher}{Society for Industrial and Applied Mathematics}, \bibinfo{address}{Philadelphia, PA}, \bibinfo{pages}{45--59}.
\newblock
\showISBNx{978-1-61197-602-1}
\urldef\tempurl%
\url{https://doi.org/10.1137/1.9781611976021.4}
\showDOI{\tempurl}


\bibitem[\protect\citeauthoryear{Kersten, Leis, Kemper, Neumann, Pavlo, and Boncz}{Kersten et~al\mbox{.}}{2018}]%
        {kersten_everything_2018}
\bibfield{author}{\bibinfo{person}{Timo Kersten}, \bibinfo{person}{Viktor Leis}, \bibinfo{person}{Alfons Kemper}, \bibinfo{person}{Thomas Neumann}, \bibinfo{person}{Andrew Pavlo}, {and} \bibinfo{person}{Peter Boncz}.} \bibinfo{year}{2018}\natexlab{}.
\newblock \showarticletitle{Everything you always wanted to know about compiled and vectorized queries but were afraid to ask}.
\newblock \bibinfo{journal}{\emph{Proceedings of the VLDB Endowment}} \bibinfo{volume}{11}, \bibinfo{number}{13} (\bibinfo{date}{Sept.} \bibinfo{year}{2018}), \bibinfo{pages}{2209--2222}.
\newblock
\showISSN{2150-8097}
\urldef\tempurl%
\url{https://doi.org/10.14778/3275366.3275370}
\showDOI{\tempurl}
\newblock
\shownote{Publisher: VLDB Endowment.}


\bibitem[\protect\citeauthoryear{Kipf, Freitag, Vorona, Boncz, Neumann, and Kemper}{Kipf et~al\mbox{.}}{2019}]%
        {kipf_estimating_2019}
\bibfield{author}{\bibinfo{person}{Andreas Kipf}, \bibinfo{person}{Michael Freitag}, \bibinfo{person}{Dimitri Vorona}, \bibinfo{person}{Peter Boncz}, \bibinfo{person}{Thomas Neumann}, {and} \bibinfo{person}{Alfons Kemper}.} \bibinfo{year}{2019}\natexlab{}.
\newblock \showarticletitle{Estimating filtered group-by queries is hard: {Deep} learning to the rescue}. \bibinfo{address}{Los Angeles, CA, USA}.
\newblock


\bibitem[\protect\citeauthoryear{Kuiper, Boncz, and Mühleisen}{Kuiper et~al\mbox{.}}{2024}]%
        {kuiper_robust_2024}
\bibfield{author}{\bibinfo{person}{Laurens Kuiper}, \bibinfo{person}{Peter Boncz}, {and} \bibinfo{person}{Hannes Mühleisen}.} \bibinfo{year}{2024}\natexlab{}.
\newblock \showarticletitle{Robust {External} {Hash} {Aggregation} in the {Solid} {State} {Age}}. In \bibinfo{booktitle}{\emph{2024 {IEEE} 40th {International} {Conference} on {Data} {Engineering} ({ICDE})}}. \bibinfo{publisher}{IEEE}, \bibinfo{address}{Utrecht, Netherlands}, \bibinfo{pages}{3753--3766}.
\newblock
\urldef\tempurl%
\url{https://doi.org/10.1109/icde60146.2024.00288}
\showDOI{\tempurl}


\bibitem[\protect\citeauthoryear{Lamb, Fuller, Varadarajan, Tran, Vandiver, Doshi, and Bear}{Lamb et~al\mbox{.}}{2012}]%
        {lamb_vertica_2012}
\bibfield{author}{\bibinfo{person}{Andrew Lamb}, \bibinfo{person}{Matt Fuller}, \bibinfo{person}{Ramakrishna Varadarajan}, \bibinfo{person}{Nga Tran}, \bibinfo{person}{Ben Vandiver}, \bibinfo{person}{Lyric Doshi}, {and} \bibinfo{person}{Chuck Bear}.} \bibinfo{year}{2012}\natexlab{}.
\newblock \showarticletitle{The vertica analytic database: {C}-store 7 years later}.
\newblock \bibinfo{journal}{\emph{Proceedings of the VLDB Endowment}} \bibinfo{volume}{5}, \bibinfo{number}{12} (\bibinfo{date}{Aug.} \bibinfo{year}{2012}), \bibinfo{pages}{1790--1801}.
\newblock
\showISSN{2150-8097}
\urldef\tempurl%
\url{https://doi.org/10.14778/2367502.2367518}
\showDOI{\tempurl}


\bibitem[\protect\citeauthoryear{Lamb, Heres, and Taylor-Davies}{Lamb et~al\mbox{.}}{2023}]%
        {lamb_aggregating_2023}
\bibfield{author}{\bibinfo{person}{Andrew Lamb}, \bibinfo{person}{Daniël Heres}, {and} \bibinfo{person}{Raphael Taylor-Davies}.} \bibinfo{year}{2023}\natexlab{}.
\newblock \bibinfo{title}{Aggregating {Millions} of {Groups} {Fast} in {Apache} {Arrow} {DataFusion} 28.0.0}.
\newblock
\newblock
\urldef\tempurl%
\url{https://arrow.apache.org/blog/2023/08/05/datafusion_fast_grouping/}
\showURL{%
\tempurl}


\bibitem[\protect\citeauthoryear{Leis, Boncz, Kemper, and Neumann}{Leis et~al\mbox{.}}{2014}]%
        {leis_morsel-driven_2014}
\bibfield{author}{\bibinfo{person}{Viktor Leis}, \bibinfo{person}{Peter Boncz}, \bibinfo{person}{Alfons Kemper}, {and} \bibinfo{person}{Thomas Neumann}.} \bibinfo{year}{2014}\natexlab{}.
\newblock \showarticletitle{Morsel-driven parallelism: a {NUMA}-aware query evaluation framework for the many-core age}. In \bibinfo{booktitle}{\emph{Proceedings of the 2014 {ACM} {SIGMOD} {International} {Conference} on {Management} of {Data}}}. \bibinfo{publisher}{ACM}, \bibinfo{address}{Snowbird Utah USA}, \bibinfo{pages}{743--754}.
\newblock
\urldef\tempurl%
\url{https://doi.org/10.1145/2588555.2610507}
\showDOI{\tempurl}


\bibitem[\protect\citeauthoryear{Li, Andersen, Kaminsky, and Freedman}{Li et~al\mbox{.}}{2014}]%
        {li_algorithmic_2014}
\bibfield{author}{\bibinfo{person}{Xiaozhou Li}, \bibinfo{person}{David~G. Andersen}, \bibinfo{person}{Michael Kaminsky}, {and} \bibinfo{person}{Michael~J. Freedman}.} \bibinfo{year}{2014}\natexlab{}.
\newblock \showarticletitle{Algorithmic improvements for fast concurrent {Cuckoo} hashing}. In \bibinfo{booktitle}{\emph{Proceedings of the {Ninth} {European} {Conference} on {Computer} {Systems}}}. \bibinfo{publisher}{ACM}, \bibinfo{address}{Amsterdam The Netherlands}, \bibinfo{pages}{1--14}.
\newblock
\urldef\tempurl%
\url{https://doi.org/10.1145/2592798.2592820}
\showDOI{\tempurl}


\bibitem[\protect\citeauthoryear{Luan and Chang}{Luan and Chang}{2022}]%
        {luan_experimental_2022}
\bibfield{author}{\bibinfo{person}{Hua Luan} {and} \bibinfo{person}{Lei Chang}.} \bibinfo{year}{2022}\natexlab{}.
\newblock \showarticletitle{An experimental study of group-by and aggregation on {CPU}-{GPU} processors}.
\newblock \bibinfo{journal}{\emph{Journal of Engineering and Applied Science}} \bibinfo{volume}{69}, \bibinfo{number}{1} (\bibinfo{date}{Dec.} \bibinfo{year}{2022}).
\newblock
\showISSN{1110-1903, 2536-9512}
\urldef\tempurl%
\url{https://doi.org/10.1186/s44147-022-00108-1}
\showDOI{\tempurl}
\newblock
\shownote{Publisher: Springer Science and Business Media LLC.}


\bibitem[\protect\citeauthoryear{Maier, Sanders, and Dementiev}{Maier et~al\mbox{.}}{2018}]%
        {maier_concurrent_2018}
\bibfield{author}{\bibinfo{person}{Tobias Maier}, \bibinfo{person}{Peter Sanders}, {and} \bibinfo{person}{Roman Dementiev}.} \bibinfo{year}{2018}\natexlab{}.
\newblock \showarticletitle{Concurrent {Hash} {Tables}: {Fast} and {General}(?)!}
\newblock \bibinfo{journal}{\emph{ACM Transactions on Parallel Computing}} \bibinfo{volume}{5}, \bibinfo{number}{4} (\bibinfo{date}{Dec.} \bibinfo{year}{2018}), \bibinfo{pages}{1--32}.
\newblock
\showISSN{2329-4949, 2329-4957}
\urldef\tempurl%
\url{https://doi.org/10.1145/3309206}
\showDOI{\tempurl}
\newblock
\shownote{Publisher: Association for Computing Machinery (ACM).}


\bibitem[\protect\citeauthoryear{Müller, Sanders, Lacurie, Lehner, and Färber}{Müller et~al\mbox{.}}{2015}]%
        {muller_cache-efficient_2015}
\bibfield{author}{\bibinfo{person}{Ingo Müller}, \bibinfo{person}{Peter Sanders}, \bibinfo{person}{Arnaud Lacurie}, \bibinfo{person}{Wolfgang Lehner}, {and} \bibinfo{person}{Franz Färber}.} \bibinfo{year}{2015}\natexlab{}.
\newblock \showarticletitle{Cache-{Efficient} {Aggregation}: {Hashing} {Is} {Sorting}}. In \bibinfo{booktitle}{\emph{Proceedings of the 2015 {ACM} {SIGMOD} {International} {Conference} on {Management} of {Data}}}. \bibinfo{publisher}{ACM}, \bibinfo{address}{Melbourne Victoria Australia}, \bibinfo{pages}{1123--1136}.
\newblock
\urldef\tempurl%
\url{https://doi.org/10.1145/2723372.2747644}
\showDOI{\tempurl}


\bibitem[\protect\citeauthoryear{Nambiar and Poess}{Nambiar and Poess}{2006}]%
        {nambiar_making_2006}
\bibfield{author}{\bibinfo{person}{Raghunath~Othayoth Nambiar} {and} \bibinfo{person}{Meikel Poess}.} \bibinfo{year}{2006}\natexlab{}.
\newblock \showarticletitle{The making of {TPC}-{DS}}. In \bibinfo{booktitle}{\emph{Proceedings of the 32nd {International} {Conference} on {Very} {Large} {Data} {Bases}}} \emph{(\bibinfo{series}{{VLDB} '06})}. \bibinfo{publisher}{VLDB Endowment}, \bibinfo{address}{Seoul, Korea}, \bibinfo{pages}{1049--1058}.
\newblock


\bibitem[\protect\citeauthoryear{Pandey, Bender, Conway, Farach-Colton, Kuszmaul, Tagliavini, and Johnson}{Pandey et~al\mbox{.}}{2023}]%
        {pandey_iceberght_2023}
\bibfield{author}{\bibinfo{person}{Prashant Pandey}, \bibinfo{person}{Michael~A. Bender}, \bibinfo{person}{Alex Conway}, \bibinfo{person}{Martin Farach-Colton}, \bibinfo{person}{William Kuszmaul}, \bibinfo{person}{Guido Tagliavini}, {and} \bibinfo{person}{Rob Johnson}.} \bibinfo{year}{2023}\natexlab{}.
\newblock \showarticletitle{{IcebergHT}: {High} {Performance} {Hash} {Tables} {Through} {Stability} and {Low} {Associativity}}.
\newblock \bibinfo{journal}{\emph{Proc. ACM Manag. Data}} \bibinfo{volume}{1}, \bibinfo{number}{1} (\bibinfo{date}{May} \bibinfo{year}{2023}), \bibinfo{pages}{1--26}.
\newblock
\urldef\tempurl%
\url{https://doi.org/10.1145/3588727}
\showDOI{\tempurl}
\newblock
\shownote{Place: New York, NY, USA Publisher: Association for Computing Machinery.}


\bibitem[\protect\citeauthoryear{Preshing}{Preshing}{2016}]%
        {preshing_leapfrog_2016}
\bibfield{author}{\bibinfo{person}{Jeff Preshing}.} \bibinfo{year}{2016}\natexlab{}.
\newblock \bibinfo{title}{Leapfrog {Probing}}.
\newblock
\newblock
\urldef\tempurl%
\url{https://preshing.com/20160314/leapfrog-probing}
\showURL{%
\tempurl}


\bibitem[\protect\citeauthoryear{Raasveldt and Mühleisen}{Raasveldt and Mühleisen}{2019}]%
        {raasveldt_duckdb_2019}
\bibfield{author}{\bibinfo{person}{Mark Raasveldt} {and} \bibinfo{person}{Hannes Mühleisen}.} \bibinfo{year}{2019}\natexlab{}.
\newblock \showarticletitle{{DuckDB}: an {Embeddable} {Analytical} {Database}}. In \bibinfo{booktitle}{\emph{Proceedings of the 2019 {International} {Conference} on {Management} of {Data}}}. \bibinfo{publisher}{ACM}, \bibinfo{address}{Amsterdam Netherlands}, \bibinfo{pages}{1981--1984}.
\newblock
\urldef\tempurl%
\url{https://doi.org/10.1145/3299869.3320212}
\showDOI{\tempurl}


\bibitem[\protect\citeauthoryear{Raman, Attaluri, Barber, Chainani, Kalmuk, KulandaiSamy, Leenstra, Lightstone, Liu, Lohman, Malkemus, Mueller, Pandis, Schiefer, Sharpe, Sidle, Storm, and Zhang}{Raman et~al\mbox{.}}{2013}]%
        {raman_db2_2013}
\bibfield{author}{\bibinfo{person}{Vijayshankar Raman}, \bibinfo{person}{Gopi Attaluri}, \bibinfo{person}{Ronald Barber}, \bibinfo{person}{Naresh Chainani}, \bibinfo{person}{David Kalmuk}, \bibinfo{person}{Vincent KulandaiSamy}, \bibinfo{person}{Jens Leenstra}, \bibinfo{person}{Sam Lightstone}, \bibinfo{person}{Shaorong Liu}, \bibinfo{person}{Guy~M. Lohman}, \bibinfo{person}{Tim Malkemus}, \bibinfo{person}{Rene Mueller}, \bibinfo{person}{Ippokratis Pandis}, \bibinfo{person}{Berni Schiefer}, \bibinfo{person}{David Sharpe}, \bibinfo{person}{Richard Sidle}, \bibinfo{person}{Adam Storm}, {and} \bibinfo{person}{Liping Zhang}.} \bibinfo{year}{2013}\natexlab{}.
\newblock \showarticletitle{{DB2} with {BLU} acceleration: so much more than just a column store}.
\newblock \bibinfo{journal}{\emph{Proceedings of the VLDB Endowment}} \bibinfo{volume}{6}, \bibinfo{number}{11} (\bibinfo{date}{Aug.} \bibinfo{year}{2013}), \bibinfo{pages}{1080--1091}.
\newblock
\showISSN{2150-8097}
\urldef\tempurl%
\url{https://doi.org/10.14778/2536222.2536233}
\showDOI{\tempurl}


\bibitem[\protect\citeauthoryear{Romanous, Windh, Absalyamov, Budhkar, Halstead, Najjar, and Tsotras}{Romanous et~al\mbox{.}}{2021}]%
        {romanous_efficient_2021}
\bibfield{author}{\bibinfo{person}{Bashar Romanous}, \bibinfo{person}{Skyler Windh}, \bibinfo{person}{Ildar Absalyamov}, \bibinfo{person}{Prerna Budhkar}, \bibinfo{person}{Robert Halstead}, \bibinfo{person}{Walid Najjar}, {and} \bibinfo{person}{Vassilis Tsotras}.} \bibinfo{year}{2021}\natexlab{}.
\newblock \showarticletitle{Efficient local locking for massively multithreaded in-memory hash-based operators}.
\newblock \bibinfo{journal}{\emph{The VLDB Journal}} \bibinfo{volume}{30}, \bibinfo{number}{3} (\bibinfo{date}{May} \bibinfo{year}{2021}), \bibinfo{pages}{333--359}.
\newblock
\showISSN{1066-8888, 0949-877X}
\urldef\tempurl%
\url{https://doi.org/10.1007/s00778-020-00642-5}
\showDOI{\tempurl}
\newblock
\shownote{Publisher: Springer Science and Business Media LLC.}


\bibitem[\protect\citeauthoryear{Vaghasiya and Jahangiri}{Vaghasiya and Jahangiri}{2024a}]%
        {vaghasiya_experiments_2024}
\bibfield{author}{\bibinfo{person}{Gaurav Vaghasiya} {and} \bibinfo{person}{Shiva Jahangiri}.} \bibinfo{year}{2024}\natexlab{a}.
\newblock \bibinfo{title}{[{Experiments} \& {Analysis}] {Hash}-{Based} vs. {Sort}-{Based} {Group}-{By}-{Aggregate}: {A} {Focused} {Empirical} {Study} [{Extended} {Version}]}.
\newblock
\newblock
\urldef\tempurl%
\url{https://doi.org/10.48550/arXiv.2411.13245}
\showDOI{\tempurl}
\newblock
\shownote{arXiv:2411.13245 [cs].}


\bibitem[\protect\citeauthoryear{Vaghasiya and Jahangiri}{Vaghasiya and Jahangiri}{2024b}]%
        {vaghasiya_hybrid_2024}
\bibfield{author}{\bibinfo{person}{Gaurav Vaghasiya} {and} \bibinfo{person}{Shiva Jahangiri}.} \bibinfo{year}{2024}\natexlab{b}.
\newblock \showarticletitle{A {Hybrid} {Approach} to {Group}-{By} and {Aggregation} {Query} {Execution}}. In \bibinfo{booktitle}{\emph{2024 {IEEE} {International} {Conference} on {Big} {Data} ({BigData})}}. \bibinfo{publisher}{IEEE}, \bibinfo{address}{Washington, DC, USA}, \bibinfo{pages}{3799--3808}.
\newblock
\showISBNx{979-8-3503-6248-0}
\urldef\tempurl%
\url{https://doi.org/10.1109/BigData62323.2024.10825803}
\showDOI{\tempurl}


\bibitem[\protect\citeauthoryear{Wejdenstål}{Wejdenstål}{[n.d.]}]%
        {wejdenstal_dashmap_nodate}
\bibfield{author}{\bibinfo{person}{Joel Wejdenstål}.} \bibinfo{year}{[n.d.]}\natexlab{}.
\newblock \bibinfo{title}{{DashMap}}.
\newblock
\newblock
\urldef\tempurl%
\url{https://github.com/xacrimon/dashmap}
\showURL{%
\tempurl}


\bibitem[\protect\citeauthoryear{Yasin}{Yasin}{2014}]%
        {yasin_top-down_2014}
\bibfield{author}{\bibinfo{person}{Ahmad Yasin}.} \bibinfo{year}{2014}\natexlab{}.
\newblock \showarticletitle{A {Top}-{Down} method for performance analysis and counters architecture}. In \bibinfo{booktitle}{\emph{2014 {IEEE} {International} {Symposium} on {Performance} {Analysis} of {Systems} and {Software} ({ISPASS})}}. \bibinfo{publisher}{IEEE}, \bibinfo{address}{CA, USA}, \bibinfo{pages}{35--44}.
\newblock
\showISBNx{978-1-4799-3606-9 978-1-4799-3604-5}
\urldef\tempurl%
\url{https://doi.org/10.1109/ISPASS.2014.6844459}
\showDOI{\tempurl}


\bibitem[\protect\citeauthoryear{Ye, Ross, and Vesdapunt}{Ye et~al\mbox{.}}{2011}]%
        {ye_scalable_2011}
\bibfield{author}{\bibinfo{person}{Yang Ye}, \bibinfo{person}{Kenneth~A. Ross}, {and} \bibinfo{person}{Norases Vesdapunt}.} \bibinfo{year}{2011}\natexlab{}.
\newblock \showarticletitle{Scalable {Aggregation} on {Multicore} {Processors}}. In \bibinfo{booktitle}{\emph{Proceedings of the {Seventh} {International} {Workshop} on {Data} {Management} on {New} {Hardware}}}. \bibinfo{publisher}{ACM}, \bibinfo{address}{Athens Greece}, \bibinfo{pages}{1--9}.
\newblock
\urldef\tempurl%
\url{https://doi.org/10.1145/1995441.1995442}
\showDOI{\tempurl}


\bibitem[\protect\citeauthoryear{Zaharia, Chowdhury, Franklin, Shenker, and Stoica}{Zaharia et~al\mbox{.}}{2010}]%
        {zaharia_spark_2010}
\bibfield{author}{\bibinfo{person}{Matei Zaharia}, \bibinfo{person}{Mosharaf Chowdhury}, \bibinfo{person}{Michael~J. Franklin}, \bibinfo{person}{Scott Shenker}, {and} \bibinfo{person}{Ion Stoica}.} \bibinfo{year}{2010}\natexlab{}.
\newblock \showarticletitle{Spark: cluster computing with working sets}. In \bibinfo{booktitle}{\emph{Proceedings of the 2nd {USENIX} {Conference} on {Hot} {Topics} in {Cloud} {Computing}}} \emph{(\bibinfo{series}{{HotCloud}'10})}. \bibinfo{publisher}{USENIX Association}, \bibinfo{address}{Boston, MA}, \bibinfo{pages}{10}.
\newblock
\urldef\tempurl%
\url{https://doi.org/10.5555/1863103.1863113}
\showDOI{\tempurl}


\end{thebibliography}

\end{document}